\documentclass[manuscript,nonacm,10pt]{acmart}
\renewcommand{\small}{\fontsize{10}{12}\selectfont}
\renewcommand{\footnotesize}{\fontsize{10}{12}\selectfont}

\makeatletter
\let\ACM@origbaselinestretch\baselinestretch
\makeatother
\usepackage{subcaption}
\usepackage{listings}
\usepackage{xcolor}
\lstdefinestyle{promptblock}{
    basicstyle=\ttfamily\footnotesize,
    breaklines=true,
    columns=fullflexible,
    keepspaces=true,
    frame=single,
    xleftmargin=0.02\textwidth,
    xrightmargin=0.02\textwidth
}
\graphicspath{{figures/}}

\setcopyright{none}
\begin{document}

\title[From Searchable to Non-Searchable]{From Searchable to Non-Searchable: Generative AI and Information Diversity in Online Information Seeking}

\author{Yulin Yu}
\affiliation{%
  \institution{Northwestern University}
  \city{Evanston}
  \state{Illinois}
  \country{USA}}

\author{Yizhou Li}
\affiliation{%
  \institution{University of Michigan}
  \city{Ann Arbor}
  \state{Michigan}
  \country{USA}}

\author{Tzu-Sheng Kuo}
\affiliation{%
  \institution{Carnegie Mellon University}
  \city{Pittsburgh}
  \state{Pennsylvania}
  \country{USA}}

\author{Kentaro Toyama}
\affiliation{%
  \institution{University of Michigan}
  \city{Ann Arbor}
  \state{Michigan}
  \country{USA}}

\author{Siddharth Suri}
\affiliation{%
  \institution{Microsoft Research}
  \city{Redmond}
  \state{Washington}
  \country{USA}}

\author{Scott Counts}
\affiliation{%
  \institution{Microsoft Research}
  \city{Redmond}
  \state{Washington}
  \country{USA}}

\begin{abstract}

Generative AI systems such as ChatGPT are transforming how people search for and engage with information online, with implications for learning, creativity, innovation, and scientific discovery. We analyze more than 200,000 real-world human-ChatGPT conversations from 2023–2025 to examine diversity in user inputs and system outputs through the lens of searchability—whether a chat conversation could be answered by a conventional search engine. We find that 79\% of user inputs are non-searchable and span broader semantic and topical spaces than searchable inputs, suggesting that generative AI expands the scope of online inquiry. However, for the same searchable prompts, Google Search returns more diverse information than ChatGPT across most topics, indicating that generative AI can narrow informational exposure even while enabling broader inquiry. Finally, the diversity of AI outputs predicts the diversity of users’ subsequent questions, revealing a feedback loop between system responses and downstream exploration.
\end{abstract}



\maketitle

\section{Introduction}

Online search capabilities and information access are fundamental to knowledge creation, scientific discovery, innovation, and social understanding. People generate new ideas, make decisions, and form beliefs by searching for and integrating information from their environments \citep{march1991exploration, pirolli1999information, kuhlthau1991inside, belkin1982ask, bates1989berrypicking, wilson1999models}. At the societal level, information diversity is a key input to creativity and innovation: exposure to heterogeneous ideas enables novel recombination, scientific breakthroughs, and cultural change \citep{merton1973sociology, weitzman1998recombinant, fleming2001recombinant, uzzi2013atypical, foster2015tradition, lee2015creativity, shi2019wisdom, jones2009burden}. The rise of the internet and search engines dramatically changed this process by reducing the cost of accessing knowledge and making large-scale information retrieval part of everyday life \citep{broder2002taxonomy, marchionini2006exploratory, brossard2013new}. Search engines now shape not only what people know, but also what questions they ask, which sources they encounter, and how they understand science, politics, health, and society \citep{rieh2002judgment, epstein2015search, delvicario2016spreading, lazer2018science, vosoughi2018spread}.

Recent advances in generative artificial intelligence (GenAI) are transforming the way people search for information online and the information they receive. Unlike traditional search engines, which return ranked lists of webpages, GenAI systems like ChatGPT synthesize information into conversational responses and support iterative, natural-language interaction \citep{radlinski2017theoretical, zamani2023conversational, webb2023emergent}. when looking from different perspectives, this shift may simultaneously expand and constrain information diversity. By lowering barriers to information seeking and enabling conversational, open-ended, and exploratory queries, GenAI may broaden the range of questions that users ask \citep{morris2010comparison, shneiderman2020human, amershi2019guidelines, gilardi2023chatgpt, argyle2023out, dellacqua2026navigating, noy2023experimental, chatterji2025people, brachman2025current}. Yet by compressing information from multiple sources into a single synthesized response, GenAI may reduce the diversity of information users encounter. This potential reduction echoes broader concerns that digital information systems can narrow exposure and homogenize knowledge consumption \citep{evans2008electronic, pariser2011filter, garrett2009echo, brundidge2010encountering, messing2014selective, bakshy2015exposure, flaxman2016filter, lorenz2011social, salganik2006experimental, cinelli2021echo, sharma2024generative}. Therefore, whether GenAI ultimately expands or narrows the information ecosystem therefore remains unclear. This question is important because information diversity is a key driver of learning, exploration, creativity, innovation, and scientific discovery \citep{march1991exploration, merton1973sociology, uzzi2013atypical, foster2015tradition, shi2019wisdom}. This motivates us to ask three research questions: 

\begin{itemize}
\item \textbf{RQ1}: Does GenAI expand the diversity of human inquiry beyond what conventional search supports?
\item \textbf{RQ2}: Does GenAI broaden or narrow the diversity of information that users receive?
\item \textbf{RQ3}: How does the diversity of GenAI responses relate to the diversity of subsequent human inquiry?
\end{itemize}



To answer these questions, we analyze 234,839 real-world human--ChatGPT conversations from 2023 to 2025 drawn from the WildChat-4.8M corpus \citep{zhao2024wildchat}. We first develop a validated computational model to identify whether a user prompt to ChatGPT is \emph{searchable} or \emph{non-searchable} through conventional web search. By searchable, we mean that information answering the prompt can be found in the top five results returned by an online search engine such as Google Search. In contrast, non-searchable prompts reflect conversational, generative, or exploratory information needs that are not readily satisfied by ranked-link retrieval alone. We then quantify the diversity of information sought by users and the diversity of information returned by AI systems through embedding-based measures. These measurements enable us to examine how GenAI reshapes information diversity on both sides of the interaction: what users ask and what systems return. We first assess whether GenAI expands the range of user inquiries beyond what traditional search supports. For prompts that remain searchable, we compare the diversity of ChatGPT responses with the diversity of information available through Google Search. Finally, we investigate whether exposure to more diverse AI-generated information predicts more diverse subsequent user inquiries.

Our analyses reveal a fundamental shift in how information is explored in the era of GenAI. Compared with conventional search engines, GenAI is associated with substantially more diverse user inputs, suggesting that users increasingly engage AI for exploratory and conversational information needs that extend beyond traditional search. At the same time, AI-generated responses are less diverse than the information returned across the top 1–10 Google search results, indicating a tendency for generative systems to consolidate information into narrower semantic representations. Longitudinal analyses of multi-turn conversations show that greater diversity in AI responses is associated with greater diversity in subsequent user questions, suggesting that AI can expand the scope of human inquiry despite producing more concentrated outputs. Together, these findings position GenAI as both an engine of exploratory information seeking and a potential bottleneck in the diversity of information that users ultimately receive.

\section{Related Work}

We first review how GenAI impacts user behavior when replacing traditional web search for information seeking. Next, we examine how these changes affect the diversity of information to which users are exposed, considering that GenAI enables interactions that extend far beyond conventional search.

\subsection{Generative AI and the Search Process}
Traditional web search is an iterative process that requires users to actively manage both the gathering and synthesis of information. Users typically search by formulating keyword queries, inspecting ranked results, opening multiple pages, and refining subsequent queries as their understanding evolves \citep{ramos2022forsense}. Sensemaking research characterizes this behavior as an interplay between two loops \citep{klein2006making,russell1993cost,dervin1992mind,pirolli1995information}. In the \textit{foraging} loop, users repeatedly seek, filter, and collect potentially useful information. In the \textit{sensemaking} loop, they organize and synthesize that material into an evolving representation of their information need \citep{kittur2013costs}. Even though ranked retrieval can lead to selective exposure \citep{adamic2005political,garrett2009echo}, the user's agency in comparing sources, reformulating queries, and synthesizing across a visible set of links preserves opportunities for serendipitous discovery \citep{dubois2018echo,guess2018avoiding}.

Conversational search powered by GenAI fundamentally reshapes the iterative process of traditional web search by automating significant parts of both the foraging and sensemaking loops\citep{radlinski2017theoretical,zamani2023conversational,yang2025searchchat}. During foraging, natural-language prompts and direct answers can reduce manual filtering and information collection across multiple webpages; comparative studies accordingly find less reliance on external webpages and less engagement with web results when conversational assistance is available \citep{kaiser2025newera,yang2025searchchat,venkit2025search}. Follow-up questions do not eliminate query reformulation but move it into the conversational context, allowing users to clarify, narrow, or expand an information need in response to the preceding answer \citep{radlinski2017theoretical,zamani2023conversational,yang2025searchchat}. During sensemaking, LLMs can organize and transform retrieved information into a coherent response, reducing some of the manual effort involved in comparing and integrating information across documents \citep{kaiser2025newera,yang2025searchchat}. This synthesis does not remove the need for sensemaking: users must still assess relevance, resolve uncertainty, and determine when additional information or verification is needed \citep{yang2025searchchat,lee2025critical}. Overall, conversational search redirect user effort from information gathering and initial synthesis toward information verification, response integration, and task stewardship \citep{lee2025critical}.

Recent work has uncovered the trade-offs stemming from this shift through several controlled studies that compare traditional and conversational interfaces across diverse contexts, ranging from everyday practical search and learning-oriented tasks to scientific inquiry \citep{kaiser2025newera,liu2024investigating,yang2025searchchat,stadler2024cognitive}. These studies show that LLM-based conversational search can achieve faster completion times and higher answer accuracy on practical tasks with lower cognitive load \citep{kaiser2025newera, stadler2024cognitive}. However, these benefits come at a cost: conversational interfaces lead users to inspect substantially fewer underlying sources with greater confirmation bias \citep{venkit2025search,kaiser2025newera,yang2025searchchat, sharma2024generative}, achieve shallower learning outcomes \citep{melumad2025experimental}, and demonstrate weaker scientific reasoning \citep{stadler2024cognitive}. A recent survey on the use of GenAI corroborates these findings, showing that greater task-specific confidence in GenAI's ability to perform a task is associated with less self-reported critical thinking effort \citep{lee2025critical}. Our work complements these controlled studies and surveys by providing a large-scale quantitative analysis of data on conversational search in the wild.

\subsection{GenAI and Information Diversity}

While research on conversational search highlights the costs of reshaping traditional foraging and sensemaking processes, GenAI introduces interaction possibilities that extend far beyond search alone. To fully understand this broader impact, recent work has examined information diversity across two distinct dimensions: the individual and the collective levels.

At the individual level, recent research suggests that GenAI can expand a user's access to information by broadening the space of accessible tasks and actionable ideas. For example, rather than merely retrieving existing facts, users can follow up on queries to request explicit explanations or conceptually reformulate ideas \citep{lichtenegger2026new, yang2025searchchat}. This expansion extends into creative work \citep{kumar2025human}. Experiments show that GenAI helps users combine remotely related concepts, resulting in higher judged creativity compared to unaided work or conventional web search \citep{lee2024empirical}. AI can also generate novel suggestions beyond the scope of existing source materials, significantly improving the novelty of short stories written by less experienced authors \citep{doshi2024generative}. Furthermore, interacting with deliberately varied AI personas can prompt different interaction dynamics that further increase diversity during collaborative ideation \citep{wan2026diverse}. In fact, AI models alone can sometimes surpass average human performance on specific divergent-thinking measures \citep{koivisto2023best}. Together, these findings demonstrate that, when considering beyond the traditional scope of search, GenAI can broaden the alternatives accessible to individual users in tasks and interface designs.

Conversely, at the collective level, research indicates that the use of GenAI can contribute to homogenization. Taking the creative task of writing as an example, studies show that writers are influenced by AI suggestions \citep{jakesch2023cowriting} and produce more predictable text \citep{arnold2020predictive}. While AI improves story quality at the individual level, stories written with AI assistance become more similar to one another across the collective scale \citep{doshi2024generative}. Similarly, a large-scale analysis of real-world college admissions essays finds that human-written texts contribute more novel semantic content to the collective pool compared to essays written with GPT-4 \citep{moon2025homogenizing}. Beyond writing, observational evidence shows that the use of GenAI in scientific research increases individual productivity but reduces the overall range of topics studied across the scientific community \citep{hao2026artificial}. Together, these findings suggest that GenAI can homogenize both the language and the ideas produced by users at a collective level.

In this work, we take a different approach to examining the effect of GenAI on exposure to diverse information. Specifically, we investigate how diversity manifests in both the user's \textit{inputs} and the resulting AI \textit{outputs}, extending our analysis beyond the scope of search processes. By leveraging real-world data in the wild, our study contributes to the body of research on how GenAI shapes information diversity through the lens of an input–output perspective.

\section{Method}

Our analysis uses 234,839 real-world human–ChatGPT conversations from 2023 to 2025, drawn from the WildChat-4.8M corpus \citep{zhao2024wildchat}. To reduce temporal imbalance in the data, we group conversations by calendar week and uniformly sample up to 2,000 conversations per week, retaining all available conversations in weeks with fewer than 2,000 conversations, which occur primarily near the beginning and end of the observation period. We define a conversation as a sequence of user and assistant turns within a single chat session. For topic-level analyses, we classify user inputs using a prompt-based taxonomy \citep{chatterji2025people}.

Our empirical analysis proceeds in three linked steps. First, we examine the information sought by users across all sampled ChatGPT conversations. We develop a validated computational measure of \emph{Input Searchability} that distinguishes between user inputs whose information needs can plausibly be satisfied by conventional web search and those that reflect conversational, generative, or exploratory needs beyond what conventional search readily provides. We then measure the diversity of information sought across these user inputs using an embedding-based measure of \emph{Input Information Diversity}.

Second, for user inputs classified as searchable, we compare the information returned by ChatGPT with that available through conventional web search. We retrieve the top 10 Google Search results for each sampled searchable user input using the Google Custom Search API\footnote{\url{https://developers.google.com/custom-search/v1/overview}} and compare their information diversity with that of the corresponding ChatGPT response. We capture this dimension using our embedding-based measure of \emph{Output Information Diversity}. Third, we examine conversational dynamics by testing whether the diversity of a ChatGPT response is associated with the diversity of the user's next inquiry. Together, these analyses allow us to examine information diversity on both sides of the interaction: the range of information users seek and the range of information returned by GenAI relative to conventional search.

Accordingly, our main analysis centers on three concepts and corresponding measures—\emph{Input Searchability}, \emph{Input Information Diversity}, and \emph{Output Information Diversity}—and then links input and output diversity across adjacent turns. We define and operationalize these components below.

\subsection{Input Searchability}

To operationalize whether a user prompt could plausibly be answered by a conventional web search engine, we construct a binary input-searchability measure. A prompt is defined as searchable if a relevant answer could be found within the top five Google Search results when the prompt is used as a query; prompts are classified as non-searchable otherwise. Non-searchable prompts include conversational, generative, creative, editing, role-play, and other open-ended requests that are not naturally handled by ranked-link retrieval.

We randomly sample 3,000 nonempty user prompts from the analysis corpus to create an annotation set. Initial labels are produced with a large language model and then verified by human annotators. Using the verified labels, we train a lightweight supervised text classifier that assigns each prompt a probability of being searchable. Let $q_i$ denote user prompt $i$, and let $s_i \in [0,1]$ denote the classifier-estimated probability that $q_i$ is searchable. We classify a prompt as searchable when $s_i \geq 0.90$:

\begin{equation}
\text{Searchable}_i =
\begin{cases}
1, & \text{if } s_i \geq 0.90, \\
0, & \text{if } s_i < 0.90.
\end{cases}
\end{equation}

We validate the classifier on a held-out sample of 100 pairs of searchable and non-searchable prompts. One human rater is asked to decide which prompt in the pair is searchable. This validation procedure yields 91\% accuracy. As an additional validation, we randomly sample 100 conversations from the corpus and manually assess whether a relevant answer can be found within the top five Google Search results for the corresponding user prompt. Comparing these manual assessments with the classifier predictions yields an accuracy of 89\% and a Cohen's kappa coefficient of 0.78, providing further evidence that the classifier reliably identifies searchable prompts. Supplementary Material Section 3 provides examples of searchable and non-searchable inputs.

\subsection{Information Diversity---Input}

To compare the diversity of user inputs based on their searchability, we operationalize input information diversity as dispersion in a semantic embedding space. Each user prompt $q_i$ is encoded into a unit-normalized embedding vector $\mathbf{x}_i$ using the SentenceTransformer model \texttt{all-MiniLM-L6-v2}. For a group of prompts $g$, such as searchable prompts, non-searchable prompts, prompts within a topic, or prompts at a given conversation turn, we compute the group centroid as

\begin{equation}
\bar{\mathbf{x}}_g = \frac{1}{N_g} \sum_{i \in g} \mathbf{x}_i,
\end{equation}
where $N_g$ is the number of prompts in group $g$. We then define each prompt's distance from the centroid as

\begin{equation}
d^{\text{in}}_{i,g} = \left\| \mathbf{x}_i - \bar{\mathbf{x}}_g \right\|_2.
\end{equation}

Following prior work on knowledge-space measurement \cite{hao2026artificial}, we define group-level input diversity as the maximum distance from the centroid:

\begin{equation}
\mathit{ID}^{\text{in}}_g = \max_{i \in g} d^{\text{in}}_{i,g}.
\end{equation}

Larger values of $\mathit{ID}^{\text{in}}_g$ indicate that the prompts in group $g$ occupy a broader semantic space. We use this measure to compare searchable and non-searchable user inputs and to examine how input diversity changes across conversation turns. To summarize the relative span of a semantic space, we also compute the range of centroid distances within each group. As robustness checks, we repeat the diversity analyses using the 95th and 90th percentiles of the distance distribution instead of the maximum.

We validate the information diversity measure using multi-turn conversations. Specifically, we select conversations containing at least five user turns and compute input information diversity for each conversation based on the semantic embeddings of its user inputs. We require at least five turns to ensure that each conversation contains a sufficient number of observations to characterize the semantic range of users' information seeking and to reduce the sensitivity of the diversity measure to any single input. This threshold provides a more stable and comparable conversation-level estimate of information diversity across conversations. We then sample conversations from the top and bottom deciles of the diversity distribution and present them side by side to two undergraduate raters. The raters are asked to identify which conversations contain broader user input content. This pairwise validation allows us to assess whether conversations assigned higher diversity scores by our measure are also perceived by human annotators as covering a broader range of information needs.
The two raters show substantial agreement (Cohen’s kappa = 0.72), and their average accuracy in identifying the high-diversity conversation is 93\%. Higher values of input information diversity indicate that users are asking about a wider range of concepts and topics, whereas lower values indicate that prompts concentrate within a narrower semantic region. We use this measure to compare the diversity of searchable and non-searchable user inputs and to examine how the breadth of user information needs evolves across conversation turns.

\subsection{Information Diversity---Output}
To measure the diversity of information returned to users, we define output information diversity using the same embedding-based framework as input information diversity, but apply it to system outputs rather than user inputs. Because our goal is to compare the diversity of information returned by GenAI with that available through conventional web search, we restrict this analysis to user inputs classified as searchable. This restriction ensures that both ChatGPT and Google Search are evaluated on information needs that can plausibly be addressed by conventional search.

For each sampled searchable user input, we retrieve the top 10 Google Search results using the Google Custom Search API\footnote{\url{https://developers.google.com/custom-search/v1/overview}} and compare these results with the ChatGPT response to the same input. Our main analysis uses the top five Google Search results, while we repeat the analysis using the top one, top three, and top 10 results as robustness checks. ChatGPT responses and Google Search results are embedded using the same SentenceTransformer model and evaluated using the output information diversity measure described below. This design allows us to compare whether ChatGPT or Google Search exposes users to a broader semantic range when responding to the same search-like information needs.

For ChatGPT, the output associated with prompt $i$ is the assistant response. For Google Search, the outputs are the returned search results for the same searchable prompt. Let $\mathbf{y}_j$ denote the unit-normalized embedding vector of output text $j$, where $j$ can refer to either a ChatGPT response or a Google Search result. For an output group $g$, we compute the centroid as

\begin{equation}
\bar{\mathbf{y}}_g = \frac{1}{M_g} \sum_{j \in g} \mathbf{y}_j,
\end{equation}

where $M_g$ is the number of output texts in group $g$. We then define each output's distance from the centroid as

\begin{equation}
d^{\text{out}}_{j,g} = \left\| \mathbf{y}_j - \bar{\mathbf{y}}_g \right\|_2.
\end{equation}

Output information diversity for group $g$ is defined as

\begin{equation}
\mathit{ID}^{\text{out}}_g = \max_{j \in g} d^{\text{out}}_{j,g}.
\end{equation}

Larger values of $\mathit{ID}^{\text{out}}_g$ indicate broader semantic coverage among the outputs returned by a system.

We further validate the embedding-based information diversity measure through a paired human evaluation. We sample 50 pairs of conversations, with each pair containing one conversation from the top 10\% of the information-diversity distribution and one conversation from the bottom 10\%. Two independent human raters are asked to identify which conversation in each pair has higher information diversity. The raters agree on 86\% of the paired comparisons, corresponding to Cohen's $\kappa = 0.72$, indicating substantial agreement.

\subsection{\texorpdfstring{{Evolution of Information Diversity in Human–AI Conversations}}{Output Comparison and Conversational Dynamics}}

To analyze conversational dynamics, we focus on adjacent user--assistant--user turn triples in longer multi-turn interactions. We construct a deterministic reservoir sample of 30,000 adjacent current-next user-turn pairs drawn from 19,148 conversations. For descriptive turn-position analyses, we retain the range with at least 30 observations per turn position. For each turn $t$, we compute the diversity of the user's current input, the diversity of the assistant's current output, and the diversity of the user's next input. In the main log-linear specification, we estimate an OLS regression in which next-turn user input diversity is predicted by current-turn AI output diversity while controlling for current-turn user input diversity:

\begin{equation}
{
\log(\mathit{ID}^{\text{in}}_{u,t+1}) = \alpha + \beta_1 z(\mathit{ID}^{\text{out}}_{a,t}) + \beta_2 z(\mathit{ID}^{\text{in}}_{u,t}) + \epsilon_t.
}
\end{equation}

Here, $\mathit{ID}^{\text{in}}_{u,t+1}$ is the diversity of the user's next-turn input, $\mathit{ID}^{\text{out}}_{a,t}$ is the diversity of the assistant output at the current turn, and $\mathit{ID}^{\text{in}}_{u,t}$ is the diversity of the user's current-turn input. The function $z(\cdot)$ denotes a standardized predictor. Robustness checks using high-percentile diversity rather than the maximum and alternative model specifications yield substantively similar conclusions.

\section{Results}

\subsection{Most ChatGPT inquiries fall outside conventional search}

We first ask whether user inquiries to ChatGPT differ from traditional search queries and whether they expand the knowledge space that users can access. Using the input-searchability measure described above, we find that across the full analysis sample, 21.0\% of prompts are classified as searchable and 79.0\% as non-searchable. This asymmetry suggests that users are not merely transferring conventional search habits to ChatGPT; instead, they are using conversational AI for requests that are less naturally handled by ranked-link retrieval.

Using the input-diversity measure, we also find that non-searchable prompts occupy a broader semantic space than searchable prompts (1.08 vs. 1.04, $P < 0.001$), which indicates that generative-AI-mediated inquiry extends across a wider range of intents and concepts. The semantic space of non-searchable prompts is 126\% larger than that of searchable prompts. Using the centroid-distance span defined in the Method section, searchable prompts span a range from 0.91 to 1.04 (range = 0.13), whereas non-searchable prompts span a range from 0.78 to 1.08 (range = 0.30). This indicates that generative-AI-mediated inquiries cover a substantially broader range of intents and concepts than prompts that can be readily expressed as traditional search queries. We also conduct robustness analyses using the 90th and 95th percentiles of the distance distribution. The results are qualitatively similar (see Figure S1 in the Supplementary Material).

\begin{figure*}[t]
\centering
\includegraphics[width=.96\linewidth]{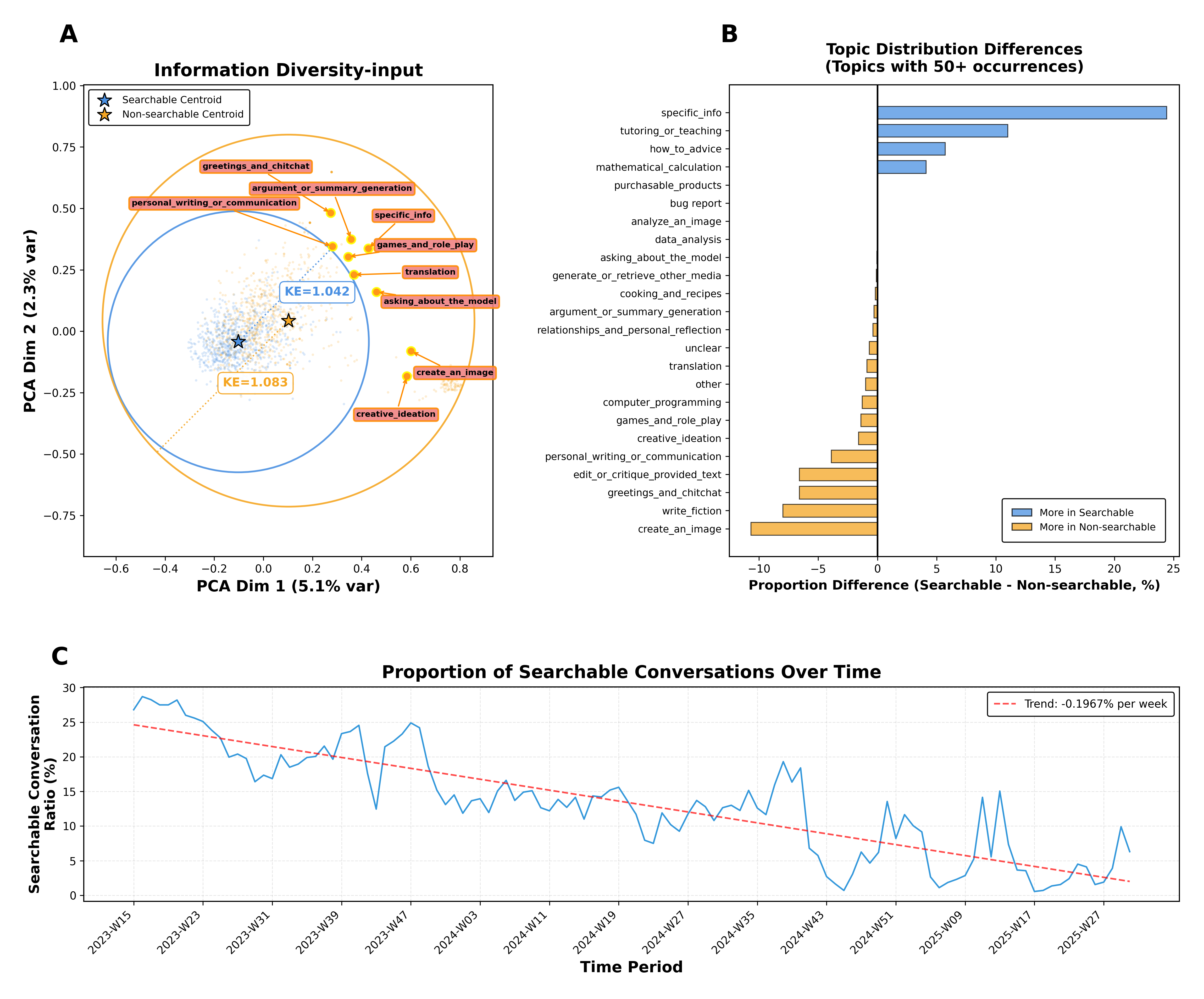}
\caption{Input diversity, topical distribution, and temporal trends in user prompt searchability. (\textbf{A}) User inputs are embedded into a 384-dimensional semantic space and group-level diversity is measured as the maximum distance from the group centroid. Non-searchable prompts occupy a broader semantic space than searchable prompts. (\textbf{B}) Topic prevalence differs markedly between searchable and non-searchable prompts. Positive values indicate topics more common in searchable prompts; negative values indicate topics more common in non-searchable prompts. (\textbf{C}) The weekly proportion of searchable conversations declines over time.}
\Description{Three panels compare searchable and non-searchable prompts. Panel A is a PCA scatter plot with blue searchable and orange non-searchable points, centroids, and extent circles; the labeled extents are 1.042 and 1.083, respectively. Panel B is a horizontal bar chart of topic-prevalence differences. Panel C is a weekly time series whose fitted trend declines from 2023 to 2025.}
\label{fig:input-diversity}
\end{figure*}

Topic differences reinforce this pattern. Searchable prompts are concentrated in domains that align with familiar search behaviors, including information retrieval, tourism, advice-seeking, and mathematical calculation. By contrast, non-searchable prompts are disproportionately associated with image generation, fiction writing, role-play, editing, and other open-ended generative tasks. Over time, the proportion of searchable prompts declines, consistent with the interpretation that users increasingly adapt to the conversational affordances of GenAI rather than treating it as a conventional search box (Fig.~\ref{fig:input-diversity}).

Figure 1C also shows that the proportion of conversations whose first user turn was searchable gradually decreased from 2023 to 2025. A weighted weekly regression confirms a significant negative coefficient (-0.0019; Table S3). The corresponding turn-level searchable-input trend is shown in Figure S2, and the first-turn conversation trend is shown in Figure S3.

\subsection{For comparable prompts, Google returns more diverse information than ChatGPT}

After examining whether GenAI expands the space of user inquiry, we next ask whether it also changes the diversity of information that users receive. Expanded inquiry does not imply expanded informational exposure: among prompts classified as searchable, Google Search returns broader semantic coverage than ChatGPT for the same user inputs.

\begin{figure*}[t]
\centering
\includegraphics[width=.96\linewidth]{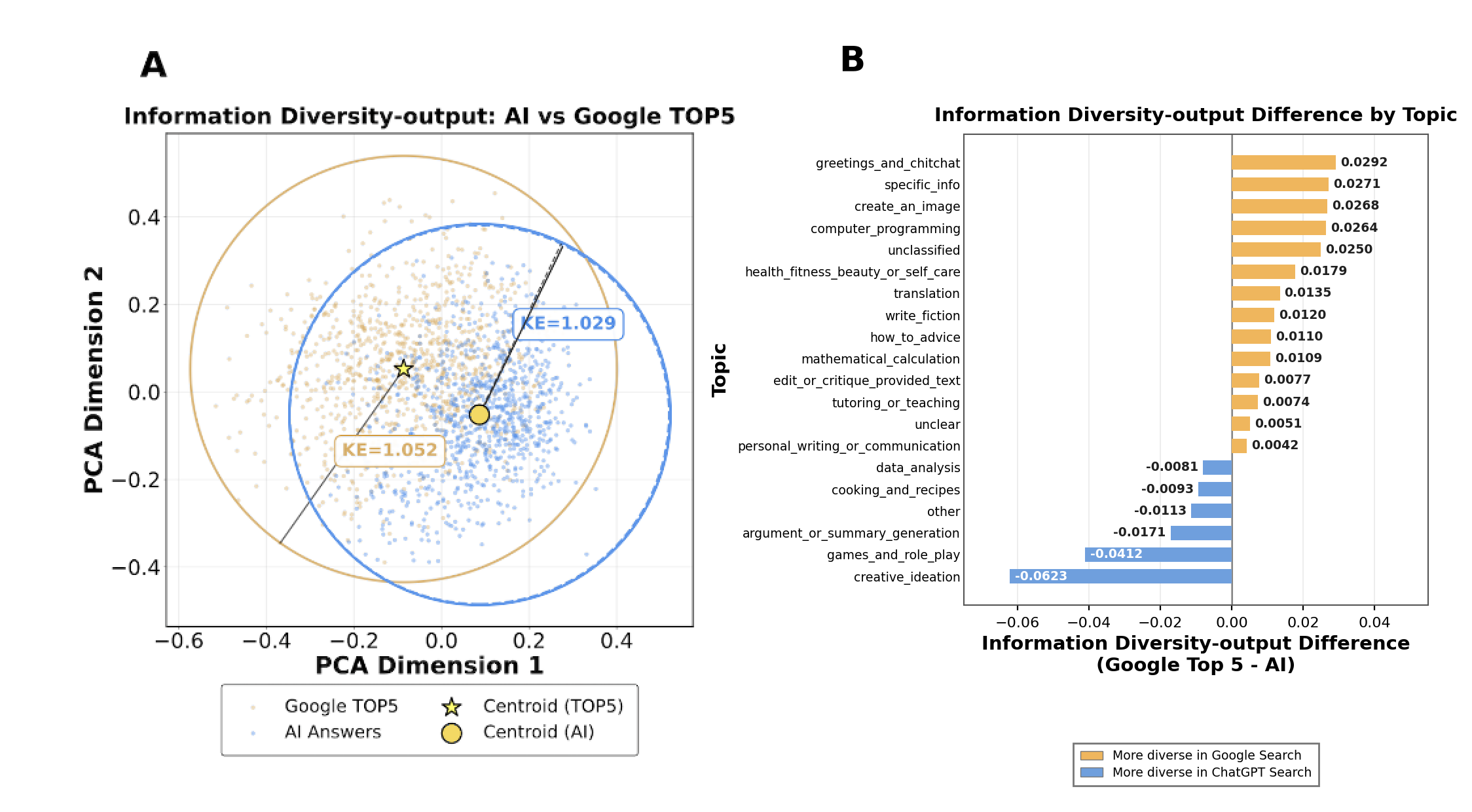}
\caption{Output diversity of ChatGPT answers versus Google Search results for the same searchable prompts. (\textbf{A}) Responses from ChatGPT and Google Search results are embedded into a shared semantic space; Google exhibits greater overall dispersion for searchable prompts. (\textbf{B}) Topic-level differences in diversity between Google top-five results and ChatGPT answers (Google minus ChatGPT). Positive values indicate broader coverage for Google; negative values indicate broader coverage for ChatGPT.}
\Description{Two panels compare the diversity of ChatGPT answers with Google Search results. Panel A is a PCA scatter plot with labeled semantic extents of 1.052 for Google's top five results and 1.029 for AI answers. Panel B ranks topic-level Google-minus-AI differences; most bars favor Google, while the largest negative bar is creative ideation, which favors AI.}
\label{fig:output-diversity}
\end{figure*}

We find that, on average, Google Search returns more diverse information than ChatGPT. Using the top five Google results, the average output diversity is 1.052 for Google versus 1.027 for ChatGPT ($P < 0.001$). The same qualitative pattern holds when we compare the top one, three, or 10 results. Topic-level analyses show that this difference is widespread and not isolated to a few domains. Most topics favor Google Search, indicating broader semantic coverage in conventional search results. One important exception is creative prompting, where ChatGPT generates more diverse outputs than Google Search. This exception is substantively meaningful: GenAI appears especially strong when users seek ideation, brainstorming, or imaginative variation, but weaker when the task involves broad informational coverage for search-like questions (Fig.~\ref{fig:output-diversity}).

\subsection{ChatGPT output diversity predicts subsequent user inquiry diversity}

We next examine whether broader AI responses are associated with broader subsequent user inquiries. The analysis focuses on adjacent user—assistant—user turn triples. For each triple, we measure the semantic diversity of the current user input, the semantic diversity of the ChatGPT response, and the semantic diversity of the next user input. We then estimate an OLS model predicting next-turn user-input diversity from current-turn ChatGPT-output diversity, controlling for current-turn user-input diversity.

Across turn positions, user inquiries become more locally focused as conversations proceed. In the reliable range retained in the Supplementary Material analysis (turn index 0--63, with at least 30 observations per turn position), high-percentile measures of user-input diversity decline relative to the first user turn: the 95th percentile diversity measure decreases by 5.38\%, and the 90th percentile measure decreases by 5.90\%. This contraction indicates that follow-up exchanges tend to narrow around the local context established earlier in the conversation (Figure S4 in the Supplementary Material).

This aggregate narrowing does not imply that AI-output diversity is irrelevant. The next-turn regression uses a deterministic reservoir sample of 30,000 adjacent turn pairs drawn from 19,148 conversations. In the main log-linear specification, a one standard-deviation increase in current-turn ChatGPT output diversity is associated with a 0.496\% increase in next-turn user-input diversity, controlling for current-turn user input diversity ($p < 0.001$). The result is robust across alternative specifications reported in Table S8 in Supplementary Material.

\begin{figure*}[t]
\centering
\includegraphics[width=.96\linewidth]{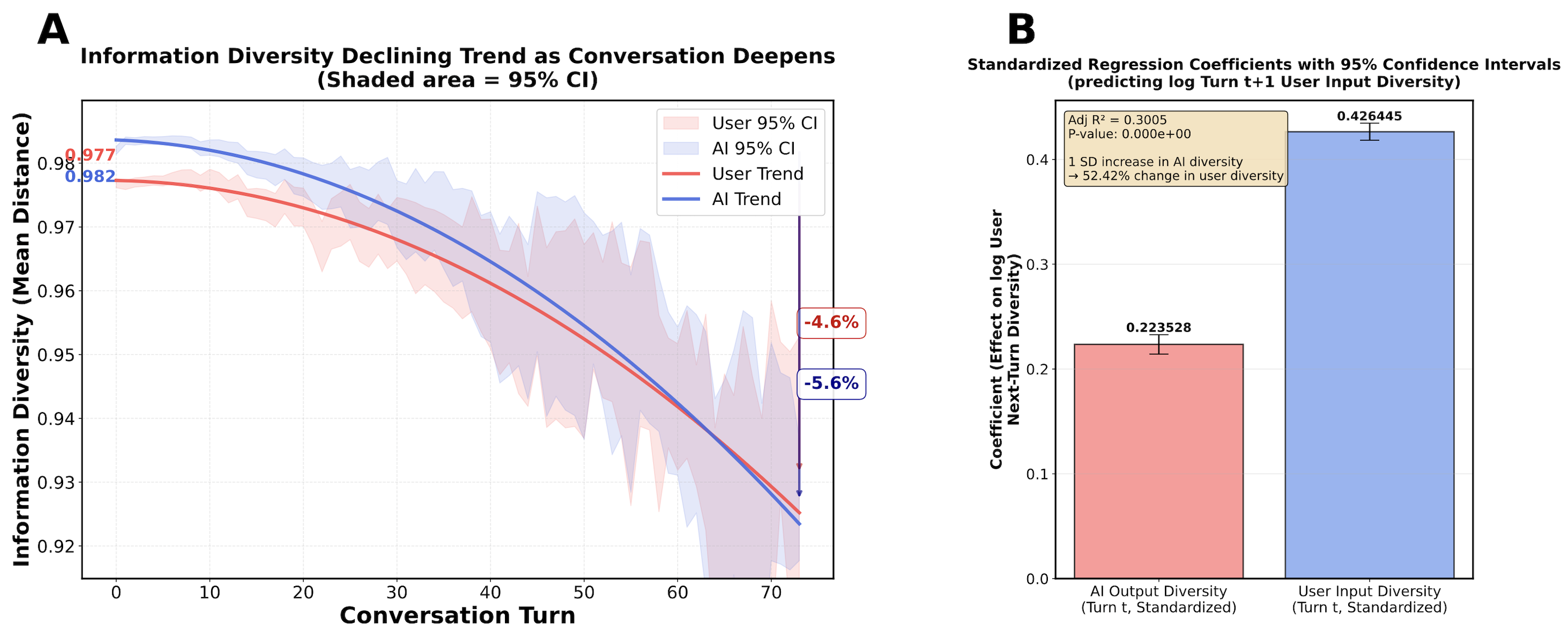}
\caption{Conversation dynamics in user input diversity. (\textbf{A}) Mean user input diversity across conversation turns in human—ChatGPT interactions. Shaded regions denote 95\% confidence intervals. (\textbf{B}) Regression coefficients from a model predicting next-turn input diversity from current-turn ChatGPT output diversity and current-turn user input diversity. Error bars denote 95\% confidence intervals.}
\Description{Panel A plots fitted user and AI information-diversity trends over conversation turns with 95 percent confidence bands; both curves decline as conversations deepen. Panel B shows standardized positive regression coefficients for current-turn AI-output diversity and current-turn user-input diversity when predicting next-turn user-input diversity, with 95 percent confidence intervals.}
\label{fig:conversation-dynamics}
\end{figure*}

Taken together, these findings suggest two simultaneous dynamics. Conversations generally become more focused over successive turns, but broader ChatGPT responses are associated with more semantically expansive follow-up prompts. This pattern is consistent with a feedback mechanism in which AI responses expose users to a wider conceptual neighborhood, increasing the diversity of subsequent inquiry even within an overall narrowing conversational trajectory (Fig.~\ref{fig:conversation-dynamics}).

\section{Discussion}

GenAI systems such as ChatGPT are transforming how people search for and engage with information online. Because information seeking underlies learning, creativity, innovation, and scientific discovery, understanding whether these systems broaden or narrow the information people encounter is important. Analyzing more than 200,000 real-world human--ChatGPT conversations from 2023 to 2025, we identify three main findings. First, 79\% of user inputs are not readily searchable through traditional search engines, and these non-searchable inputs span a broader semantic and topical space than searchable inputs, suggesting that GenAI expands the space of inquiry beyond traditional search. Second, for prompts that remain comparable to traditional search, Google Search returns more diverse information than ChatGPT across most topics, indicating that broader inquiry on the input side does not necessarily translate into broader informational exposure on the output side. Finally, greater diversity in ChatGPT responses predicts greater diversity in users' subsequent questions, suggesting a feedback relationship between the information generated by AI and downstream user exploration. Together, these findings reveal a central tension in generative information seeking: conversational AI expands the range of questions people ask, while its responses can compress the information users encounter, and the diversity of those responses may in turn shape subsequent inquiry.
\subsection{Broadening Inquiry While Narrowing Information Exposure}

Our findings reveal a central tension in GenAI search. On the one hand, ChatGPT expands the space of inquiry: the large majority of prompts in our corpus are not easily reducible to conventional search-engine queries, and these prompts span a broader topical and semantic range than do searchable inputs. On the other hand, when users ask questions that remain comparable to traditional search, ChatGPT often returns less diverse information than Google. GenAI therefore broadens what people ask while potentially narrowing what they see. This pattern is consistent with evidence that digital systems can concentrate attention and exposure, and our results show the same trade-off in naturally occurring human-AI conversations \citep{salganik2006experimental, lorenz2011social, sharma2024generative}.

This tension connects two perspectives that have largely been studied separately. Studies of conversational search show that GenAI changes task efficiency, source engagement, cognitive load, and learning \citep{kaiser2025newera,liu2024investigating,yang2025searchchat,stadler2024cognitive}, while work on AI-assisted creativity and homogenization has examined how AI affects the diversity of downstream artifacts and topic portfolios \citep{lee2024empirical,doshi2024generative,moon2025homogenizing,hao2026artificial}. Our findings suggest that these processes may begin earlier, within the information-seeking interaction itself. By separating the breadth of users' requests from the breadth of returned content, we show that naturalistic ChatGPT use can simultaneously expand the space of inquiry and compress the information returned for comparable searchable prompts. Moreover, the positive association between response distinctiveness and the distinctiveness of the subsequent prompt suggests that these two sides of the interaction may be dynamically linked. 

\subsection{Information Compression and the Changing Search Process}

This tension should not be interpreted as implying that lower output diversity is always undesirable. In many fact-seeking contexts, users may prefer a concise, synthesized answer over a diverse list of sources. This interpretation is consistent with information foraging and information-behavior theories, which emphasize that users seek information under constraints of effort, uncertainty, time, and expected value~\cite{pirolli1999information, belkin1982ask, wilson1999models}. Compared with traditional search engines, which typically expose users to multiple ranked results, ChatGPT often collapses information into a single generated response. This can reduce search costs and help users quickly identify an answer, especially when the information need is well defined. Our output-diversity result likely reflects this compression: synthesis lowers comparison effort but collapses source-level variation into one response. In this sense, ChatGPT may function less like a conventional search engine and more like an answer engine: it synthesizes, explains, translates, transforms, and adapts information in ways that traditional keyword search does not easily support \citep{radlinski2017theoretical}.

This compression perspective also helps connect our findings to theories of information foraging and sensemaking. Traditional search requires users to select, compare, and integrate information across visible sources \citep{russell1993cost,kittur2013costs,ramos2022forsense}, whereas conversational systems can shift part of this work to the model. Consistent with this shift, controlled studies find fewer visits to external webpages and less engagement with search results in conversational settings \citep{kaiser2025newera,yang2025searchchat,venkit2025search}, while users increasingly face the task of verifying and integrating generated outputs \citep{lee2025critical}. Although our data do not measure source inspection or cognitive effort directly, the narrower semantic space of ChatGPT responses provides complementary evidence about what users receive when this synthesis is delegated to AI. Information compression may therefore be one informational consequence of reducing the need to navigate and compare multiple sources.

\subsection{When Does Information Diversity Matter?}

At the same time, the value of diversity depends strongly on the task. For narrow factual questions, diversity may be less important than accuracy, clarity, and source reliability. For exploratory search, learning, brainstorming, creativity, and sensemaking, however, diversity is central. Prior work on search-result diversification shows that information retrieval systems must balance relevance with novelty, as redundant results may satisfy narrow queries but limit exposure to alternative information \citep{carbonell1998use}. Research on curiosity and the berrypicking model of information seeking—which posits that information needs evolve through successive queries, partial answers, and unexpected cues—suggests that information needs evolve dynamically over the course of search \citep{bates1989berrypicking, loewenstein1994psychology, ingwersen1996cognitive}. Similarly, theories of exploration and exploitation suggest that users and organizations must balance the refinement of known information with the search for new possibilities \citep{march1991exploration}. From this perspective, GenAI may be most valuable when treated as a flexible information tool rather than a direct replacement for Google Search. By default, ChatGPT may provide a relatively narrow synthesized answer, but when users explicitly ask it to brainstorm, compare perspectives, generate alternatives, or explore a topic broadly, it may produce a more diverse and useful response space.

This distinction is especially important because the same reduction in diversity may have different consequences across tasks. Conversational AI can reduce source engagement and perceived cognitive effort, but these efficiencies can sometimes coincide with weaker reasoning or critical engagement \citep{kaiser2025newera,yang2025searchchat,stadler2024cognitive,lee2025critical}. Our results add another dimension to this trade-off: the semantic breadth of the returned information itself. Google covered a broader semantic range for most searchable topics, whereas ChatGPT was more diverse for creative prompting. This heterogeneity reinforces the idea that diversity should not be treated as uniformly beneficial. Because we do not measure accuracy, task success, or user utility, our results instead suggest a task-dependent hypothesis: semantic compression may be relatively useful for well-defined factual needs but more consequential when users are exploring, learning, or searching for unfamiliar possibilities.

\subsection{Information Exposure, Creativity, and Homogenization}

This task-dependent interpretation also connects our findings to research on creativity and innovation. A long line of work suggests that novel ideas often emerge from recombining distant or atypical knowledge domains \citep{weitzman1998recombinant, fleming2001recombinant, uzzi2013atypical, lee2015creativity}. If conversational AI systems provide narrower information spaces by default, they may reduce users' exposure to the diverse materials from which new ideas emerge. This concern is consistent with recent evidence that GenAI can improve some individual creative outcomes while altering the diversity and originality of generated content \citep{doshi2024generative, lee2024empirical, koivisto2023best, kumar2025human}. Thus, ChatGPT may simultaneously help individual users produce useful outputs while making the broader information environment more homogeneous.

Our findings also point to a possible upstream source of the homogenization documented in creative and knowledge-production settings. Existing studies have primarily measured convergence in downstream outputs, such as stories, essays, or scientific topic portfolios \citep{doshi2024generative,moon2025homogenizing,hao2026artificial}. We instead examine the information returned before such outputs are produced. ChatGPT's more concentrated system-level response space, together with the positive association between response distinctiveness and subsequent prompt distinctiveness, suggests that the structure of information exposure may matter for later exploration or convergence. We do not observe which information users actually attend to or incorporate, nor do we connect responses directly to downstream creative artifacts, so this should be interpreted as a plausible pathway rather than an established mechanism. 
The creative-ideation exception further suggests that conversational AI can expand rather than compress the response space when it is used for brainstorming, alternative generation, and open-ended exploration. Evidence that deliberately diversified AI inputs can preserve collective creative diversity \citep{wan2026diverse} further motivates treating returned-content diversity as an important design variable whose downstream consequences should be tested directly.

\subsection{GenAI Beyond Search}

These results also clarify how the transition from search engines to GenAI is uneven across tasks. Searchable prompts remain common enough to matter, but they no longer dominate the interaction landscape. Instead, users appear to be adopting conversational systems for synthesis, transformation, creativity, and exploratory reasoning. For these use cases, ChatGPT may provide capabilities that traditional search engines do not: it can reframe a question, generate candidate ideas, compare alternatives, and support iterative refinement through conversation. This is consistent with evidence that GenAI can improve productivity in writing and knowledge-work tasks \citep{noy2023experimental, dellacqua2026navigating, brachman2025current}. However, our findings suggest that these advantages may come with a homogenizing tendency, especially when users rely on default responses rather than explicitly prompting for breadth.

The prevalence of non-searchable prompts further suggests that this transition is broader than a substitution between two retrieval interfaces. Users have been shown to turn to conversational systems for explanations, reformulations, examples, and guidance-oriented or output-oriented tasks beyond factual retrieval \citep{lichtenegger2026new,yang2025searchchat}. Our large-scale observational data show how prominent these uses have become: 79\% of prompts were classified as non-searchable under our operational definition, with many concentrated in editing, creative generation, role-play, and other open-ended activities. This does not imply that conventional search could never support such tasks. Rather, it indicates that much of observed ChatGPT use is not naturally represented as a request for a ranked set of retrievable webpages, underscoring the broader transformation of information-seeking behavior enabled by conversational AI.

\subsection{Design Implications for Generative Search}

The practical implication is that future generative search systems should support different modes of information access. For factual and lookup-oriented tasks, systems should prioritize accuracy, source grounding, transparency, and verification \citep{rieh2002judgment, shneiderman2020human}. For exploratory and creative tasks, systems should make diversity an explicit design goal by surfacing multiple perspectives, alternative explanations, disagreement among sources, and conceptually distant possibilities. Hybrid search systems that combine generated answers with diverse evidence may be especially important for preserving broad exposure when the task is informational, while still leveraging the efficiency and flexibility of conversational AI \citep{yen2024search}.

Taken together, these patterns suggest that generative search systems may benefit from adapting the form of information access to the user's task and underlying motivation. Future systems could first infer whether a user is primarily seeking a specific answer or would benefit from broader exploration, brainstorming, or sensemaking, and then adapt their response strategy accordingly. For answer-seeking users, systems could prioritize concise synthesis, accuracy, and inspectable evidence; for exploration-oriented users, they could deliberately surface diverse perspectives, alternative framings, and conceptually distant possibilities rather than prematurely converging on a single answer. Because such motivations may be difficult to infer reliably and may evolve over the course of an interaction, systems could also provide explicit controls that allow users to shift between more convergent and more exploratory modes—for example, between obtaining a focused answer and exploring a broader set of possibilities. Hybrid interfaces already illustrate how generated synthesis can be combined with inspectable search evidence, while other work shows that deliberately varying AI-provided ideas can preserve diversity \citep{yang2025searchchat,yen2024search,wan2026diverse}. Our results provide empirical motivation for such task-contingent designs: searchable and non-searchable prompts occupy different semantic spaces, the ChatGPT vs. Google diversity gap varies across topics, and response distinctiveness is associated with the distinctiveness of subsequent inquiry. We do not directly test interface interventions or user utility, so these findings motivate rather than validate a specific design. A promising direction is therefore to design generative search systems that dynamically balance convergence and exploration based on users' informational goals, while giving users explicit control over the breadth and diversity of the information they encounter.

\subsection{Limitations and Future Work}

The study has several limitations. We do not observe a parallel corpus of real-world search engine interactions from the same users and time period, so our comparison relies on Google results retrieved for prompts sampled from ChatGPT interactions. In addition, our analysis relies on the WildChat dataset, which may introduce selection bias in the users and interactions represented in our sample. WildChat was collected by providing free access to ChatGPT through a third-party interface to users who consented to share their conversations \citep{zhao2024wildchat}. As a result, its users may differ systematically from the broader population of ChatGPT users in their demographics, technical familiarity, motivations, or patterns of use. The distribution of tasks and interaction behaviors observed in WildChat should therefore not be interpreted as representative of all GenAI use, and the prevalence and diversity of searchable and non-searchable interactions may differ across platforms, user populations, and time periods. Future work could test the generalizability of these patterns using interaction logs from other GenAI systems and more representative user populations. In addition, our measures capture semantic dispersion rather than factual correctness, epistemic quality, or usefulness. A narrower ChatGPT answer may sometimes be more accurate or more useful than a broader set of Google results, while a more diverse answer may also include irrelevant or unreliable information. This distinction is especially important because large language models can generate fluent but unsupported or inaccurate content, making factuality and verifiability central concerns for generative search systems \citep{ji2023survey,gao2023retrieval}. Future work should therefore evaluate diversity together with factual accuracy, source quality, user trust, and decision quality. In particular, future studies should compare whether ChatGPT and Google differ not only in the breadth of information they provide, but also in whether users reach correct answers, identify uncertainty, and make better decisions.

More broadly, our findings suggest that GenAI might change the structure of information seeking. It expands the kinds of questions people ask, but its default mode of response may compress the information environment into a narrower synthesized view. Whether this compression is beneficial or harmful depends on the user's goal. For finding an answer, it may improve efficiency. For brainstorming, learning, and innovation, it may limit exposure to the diverse materials from which new ideas emerge.

\bibliographystyle{ACM-Reference-Format}
\bibliography{references}

\clearpage
\begingroup
\raggedbottom
\setcounter{section}{0}
\setcounter{figure}{0}
\setcounter{table}{0}
\renewcommand{\thesection}{S\arabic{section}}
\renewcommand{\thefigure}{S\arabic{figure}}
\renewcommand{\thetable}{S\arabic{table}}
\renewcommand{\theHsection}{supp.\arabic{section}}
\renewcommand{\theHfigure}{supp.\arabic{figure}}
\renewcommand{\theHtable}{supp.\arabic{table}}
\def\shorttitle{Supplementary Material}
\setlength{\intextsep}{12pt}
\setlength{\textfloatsep}{14pt}
\setlength{\floatsep}{14pt}
\captionsetup{font=small,labelfont=bf,justification=raggedright,singlelinecheck=false}
\captionsetup[subfigure]{font=footnotesize,skip=5pt}
\setlength{\parskip}{4pt plus 1pt}

\begin{center}
{\sffamily\bfseries\fontsize{22}{26}\selectfont Supplementary Material\par}
\vspace{8pt}
{\sffamily\large From Searchable to Non-Searchable:\\
GenAI and Information Diversity in Online Information Seeking\par}
\end{center}
\vspace{4pt}
\noindent{\color{black!25}\rule{\linewidth}{0.5pt}}
\vspace{8pt}

\noindent{\sffamily\bfseries Contents}\par
\begingroup
\small
\setlength{\tabcolsep}{0pt}
\renewcommand{\arraystretch}{1.45}
\noindent\begin{tabular*}{\linewidth}{@{\extracolsep{\fill}}lr}
\hyperref[sec:si-data]{S1\quad Data description} & \pageref*{sec:si-data} \\
\hyperref[sec:si-figures]{S2\quad Supplementary figures and robustness checks} & \pageref*{sec:si-figures} \\
\hyperref[sec:si-examples]{S3\quad Examples of searchability and input--output diversity} & \pageref*{sec:si-examples} \\
\hyperref[sec:si-prompts]{S4\quad Searchability classification prompt} & \pageref*{sec:si-prompts} \\
\hyperref[sec:si-tables]{S5\quad Supplementary tables} & \pageref*{sec:si-tables} \\
\end{tabular*}
\endgroup
\vspace{12pt}

\section{Data Description}
\label{sec:si-data}

Our analysis is based on a corpus of 234,839 real-world conversations between humans and ChatGPT collected from 2023 to 2025, drawn from the \textsc{WildChat-4.8M} dataset\cite{zhao2024wildchat}\footnote{\url{https://huggingface.co/datasets/allenai/WildChat-4.8M}}
. This dataset is a large-scale collection of real user-ChatGPT interaction logs, where users interacted with ChatGPT via public interfaces operated by the Allen Institute. To mitigate temporal imbalance and manage computational constraints, we construct a temporally balanced analysis subset. Specifically, we group conversations by calendar week (based on UTC timestamps) and uniformly sample up to 2,000 English conversations per week. In this dataset, a conversation is defined as a sequence of user-assistant message turns occurring within a single chat session.
To enable topic-based analysis, we also label user inputs using the prompt-based taxonomy \cite{chatterji2025people}.

\clearpage
\section{Supplementary Figures and Robustness Checks}
\label{sec:si-figures}

\begin{figure}[H]
    \centering

    \begin{subfigure}[t]{0.82\textwidth}
        \centering
        \includegraphics[width=\textwidth]{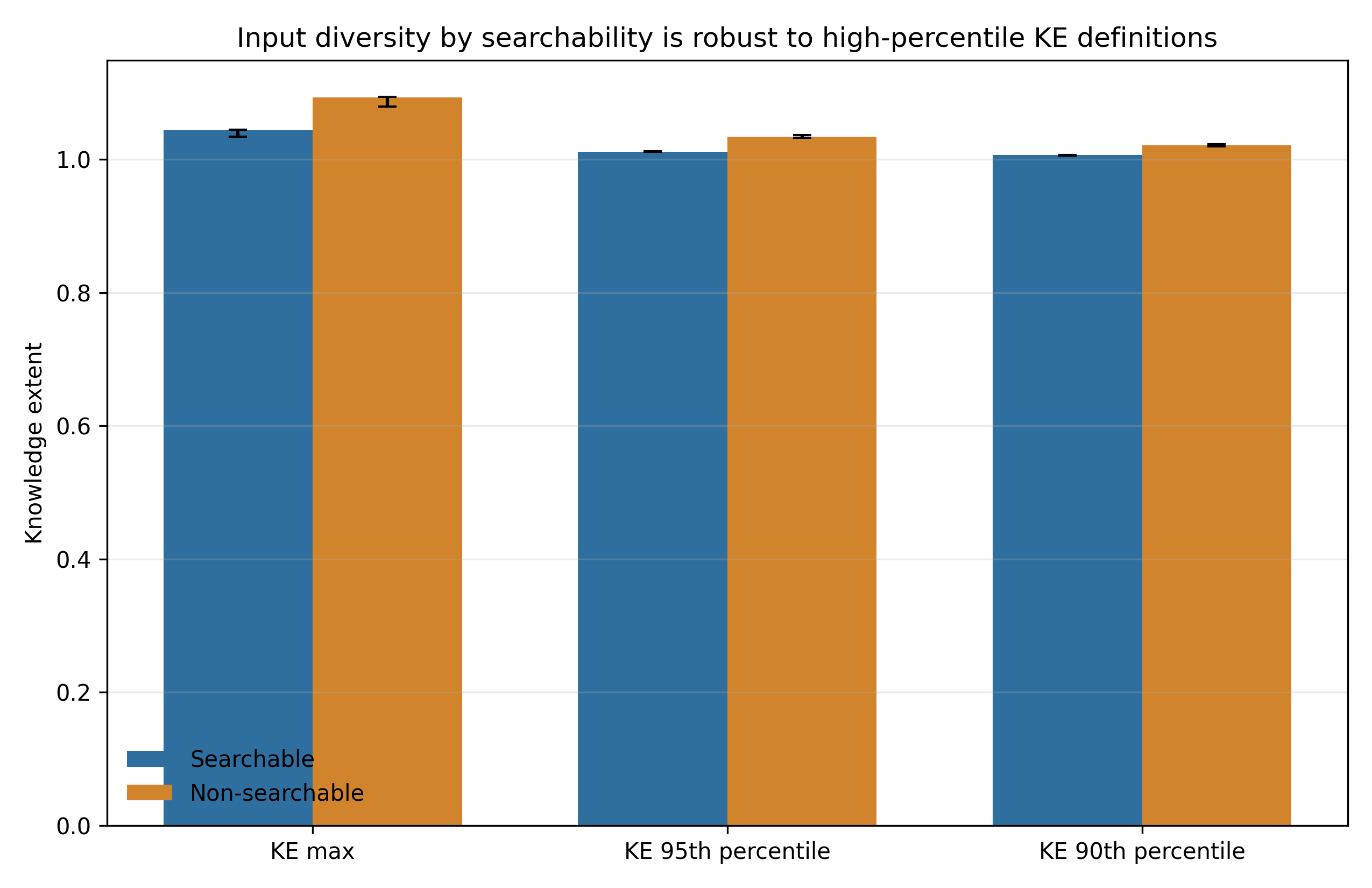}
        \caption{Searchable and non-searchable inputs show a consistent difference in knowledge extent across KE max, KE 95th percentile, and KE 90th percentile definitions.}
        \label{fig:input_diversity_searchability_bar}
    \end{subfigure}

    \vspace{0.4em}

    \begin{subfigure}[t]{0.43\textwidth}
        \centering
        \includegraphics[width=\textwidth]{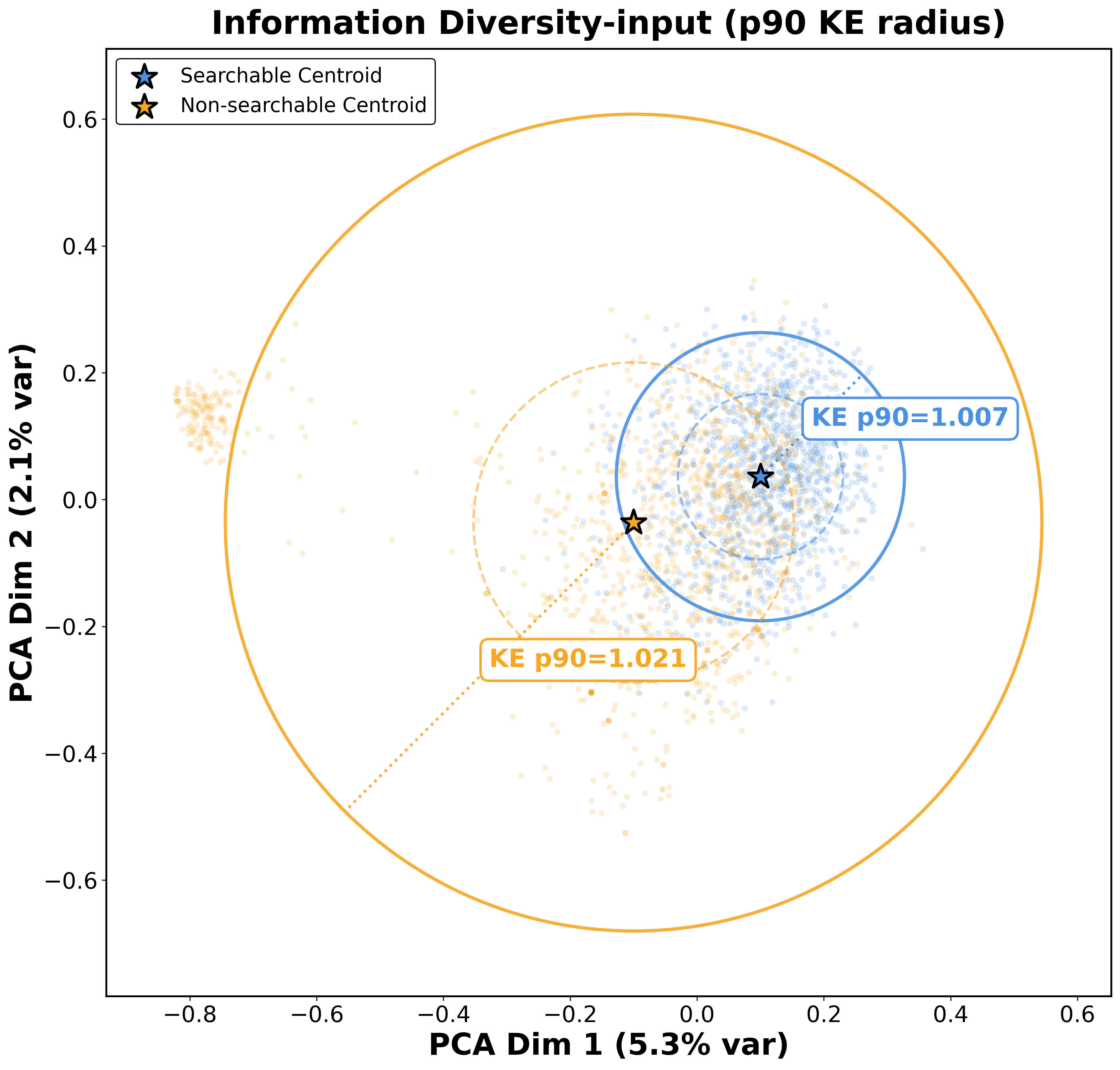}
        \caption{Input diversity measured using the KE 90th percentile radius.}
        \label{fig:input_diversity_circle_p90}
    \end{subfigure}
    \hfill
    \begin{subfigure}[t]{0.43\textwidth}
        \centering
        \includegraphics[width=\textwidth]{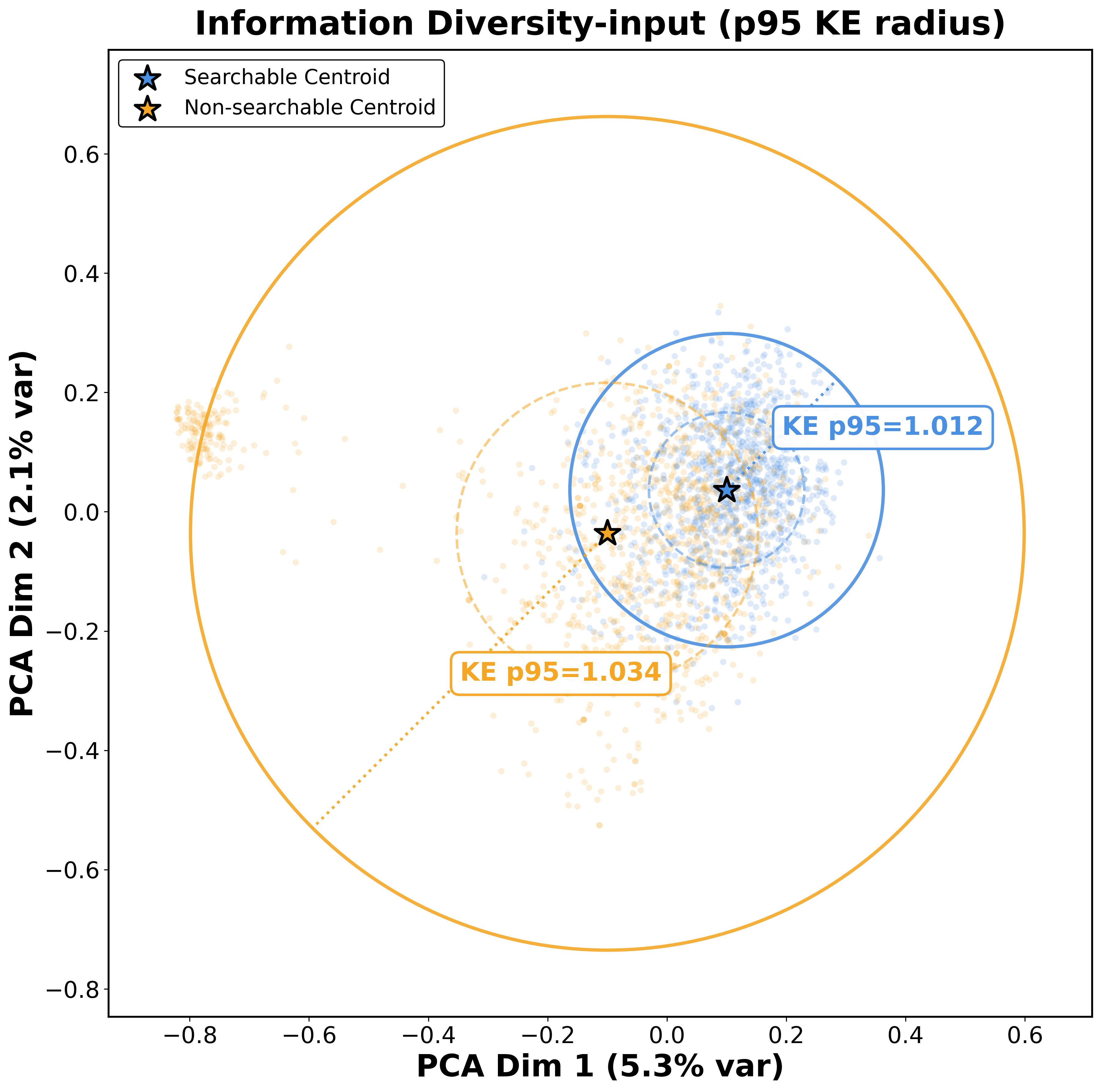}
        \caption{Input diversity measured using the KE 95th percentile radius.}
        \label{fig:input_diversity_circle_p95}
    \end{subfigure}

    \caption{Input diversity by searchability is robust to high-percentile definitions of knowledge extent. The aggregate comparison remains consistent across KE max, KE 95th percentile, and KE 90th percentile measures, and the PCA visualizations show that non-searchable inputs occupy a broader semantic region than searchable inputs under both p90 and p95 KE radius definitions.}
    \label{fig:input_diversity_high_percentile_robustness}
    \Description{A bar chart compares searchable and non-searchable input knowledge extent under max, p95 and p90 definitions. Two PCA plots show the corresponding p90 and p95 input regions.}
\end{figure}

\clearpage
\subsection{Weekly searchability trends}
\begin{figure}[H]
    \centering
    \includegraphics[width=0.96\textwidth]{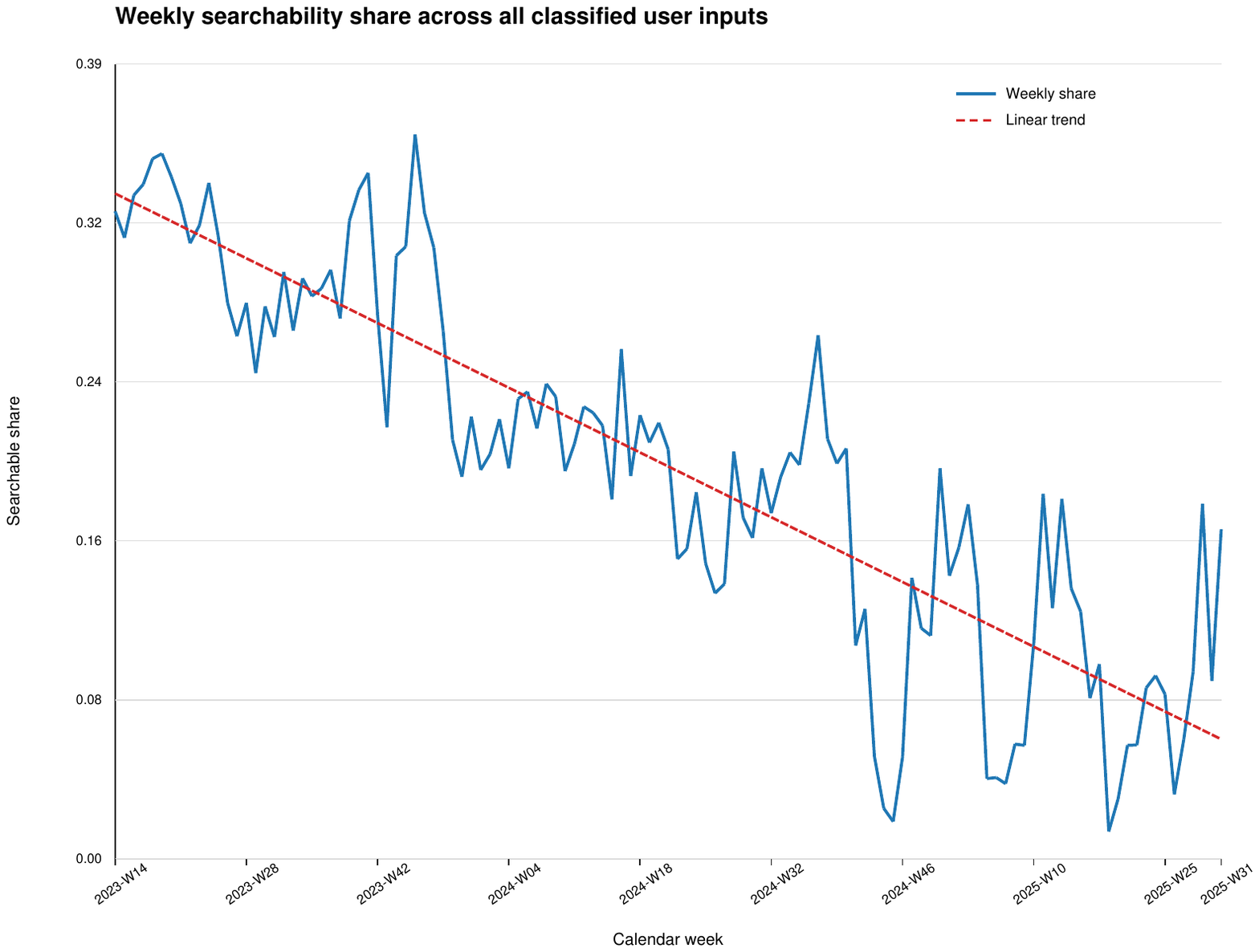}
    \caption{Weekly share of classified user inputs labeled searchable. The line plot shows the weekly searchable share across all classified user inputs, with a fitted linear trend.}
    \label{fig:weekly_turn_searchability}
    \Description{A weekly line plot of the searchable share of classified user inputs, with a fitted downward linear trend.}
\end{figure}

\begin{figure}[H]
    \centering
    \includegraphics[width=0.96\textwidth]{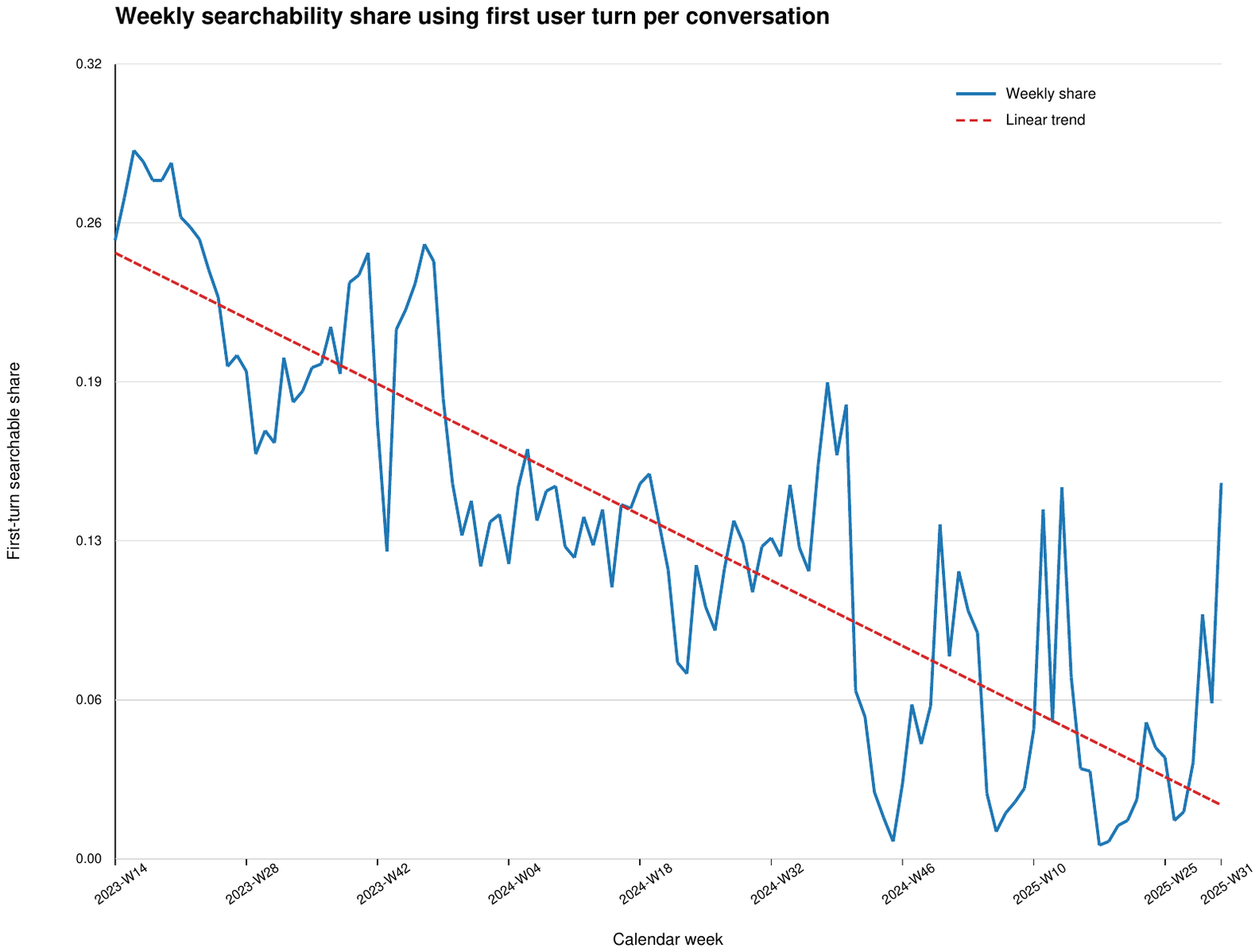}
    \caption{Weekly share of conversations whose first user turn is labeled searchable. The line plot shows the first-turn searchable conversation share by calendar week, with a fitted linear trend.}
    \label{fig:weekly_first_turn_searchability}
    \Description{A weekly line plot of the share of conversations with a searchable first user turn, with a fitted downward linear trend.}
\end{figure}

\clearpage
\subsection{Turn-position contraction is robust to KE max, p95 and p90}
\begin{figure}[H]
    \centering
    \includegraphics[width=0.95\textwidth]{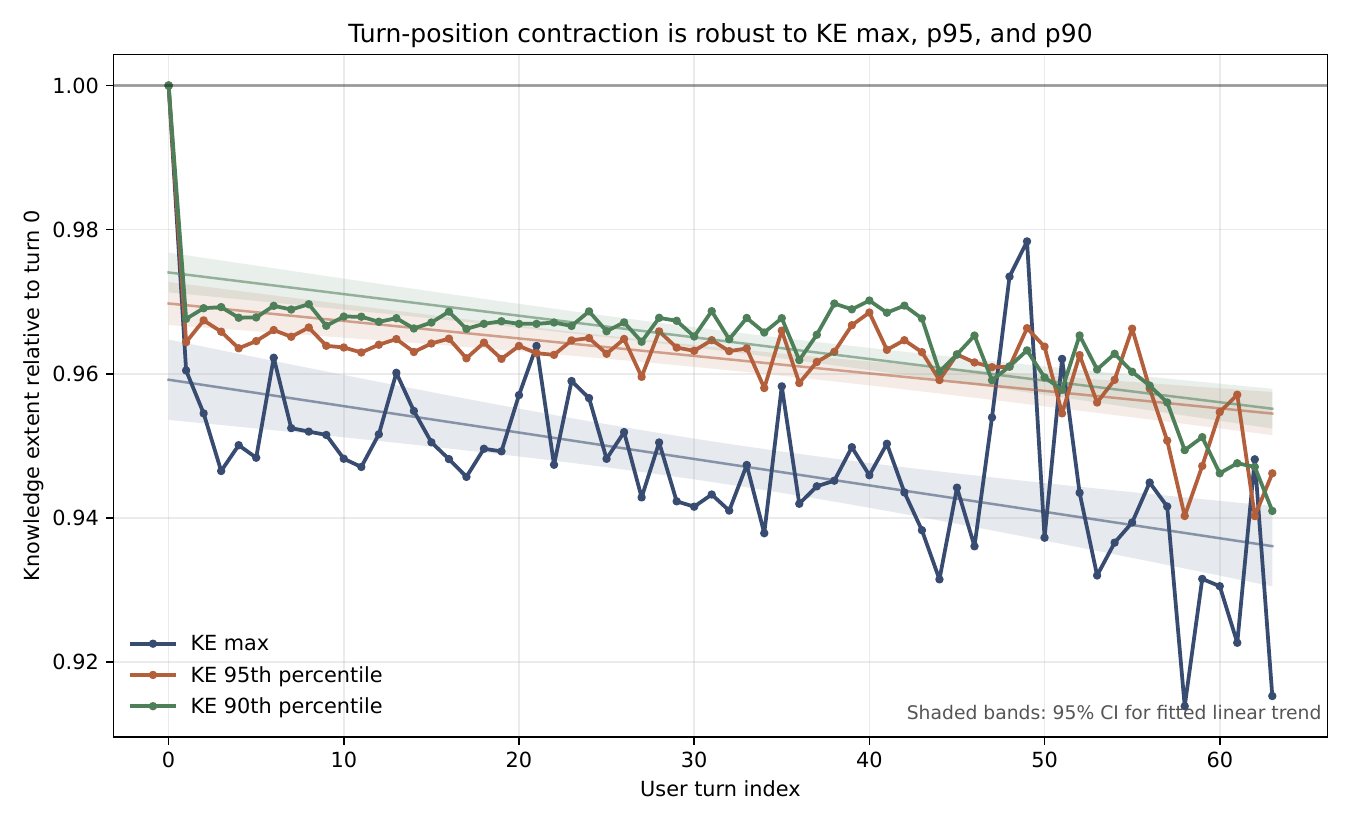}
    \caption{Turn-position contraction is robust to alternative high-percentile definitions of knowledge extent. Knowledge extent is normalized relative to turn 0 and computed using KE max, KE 95th percentile, and KE 90th percentile.}
    \label{fig:ke_percentile_turn_position_robustness}
    \Description{Three normalized knowledge-extent curves, using max, p95 and p90, across user-turn positions, with fitted linear trends and 95 percent confidence bands.}
\end{figure}

\clearpage
\section{Examples of Searchable and Non-Searchable Queries with High and Low Input-Output Diversity}
\label{sec:si-examples}
\begingroup
\setstretch{1}
\setlength{\parskip}{2pt}

The following examples illustrate the constructs used in the analyses. Candidate examples were selected from the classified conversation corpus using either classifier labels or embedding distance from the relevant centroid, and then manually screened for interpretability. We removed ambiguous fragments, context-dependent outputs, persona artifacts, and rows that did not clearly illustrate the intended concept. URLs, email addresses, phone-like strings, and long numeric identifiers were replaced with bracketed placeholders; long excerpts were truncated for readability.

\subsection*{High input diversity}
These inquiries are in the upper tail of distance from the input-space centroid. They illustrate prompts that are semantically or structurally far from the average user inquiry.

\begin{itemize}
\setlength{\itemsep}{3pt}
\setlength{\parsep}{0pt}
\setlength{\parskip}{0pt}
\interlinepenalty=10000
\item \emph{I want to control auto-redirect per request using the same HttpClient. But this is not possible because if you tried to change AllowAutoRedirect it throws an exception: This instance has already started one or more requests. Properties can only be modified be...} (distance percentile=100.0). Combines a specific programming framework, exception text, and implementation constraint, placing it in the input-space tail.
\item \emph{please check incorrect answers. How does React Router manage scroll restoration during navigation? React Router uses the browser's default scroll restoration behavior with client-side routing, ensuring the scroll position is restored to the necessary spot whe...} (distance percentile=100.0). Embeds a candidate technical answer and asks the model to verify it, making the prompt more specialized than a typical request.
\item \emph{140 investment banks send 3,285 FedEx packages a month. The average price of a package is 12.5 and the cost is 4.25. FedEx's yearly retention rate among investment banks is 80\%. On average, FedEx spends \$7,000 to acquire a new investment bank. Assume a yearly...} (distance percentile=100.0). Contains a multi-step quantitative business calculation with several numeric assumptions.
\end{itemize}

\subsection*{Low input diversity}
These inquiries are close to the input-space centroid. They illustrate common, compact requests such as editing, rewriting, or interface help.

\begin{itemize}
\setlength{\itemsep}{3pt}
\setlength{\parsep}{0pt}
\setlength{\parskip}{0pt}
\interlinepenalty=10000
\item \emph{How can I print this conversation?} (distance percentile=0.0). A short, generic interface/help request close to common user-input patterns.
\item \emph{5 title proposals for this text} (distance percentile=0.0). A compact rewriting request asking for title proposals from supplied text.
\item \emph{ok but can you write it with simpler english} (distance percentile=0.0). A common simplification/editing request that stays near the input centroid.
\end{itemize}

\subsection*{High output diversity}
These assistant responses are in the upper tail of distance from the output-space centroid. They illustrate specialized answer domains in the response space.

\begin{itemize}
\setlength{\itemsep}{3pt}
\setlength{\parsep}{0pt}
\setlength{\parskip}{0pt}
\interlinepenalty=10000
\item \emph{In an ASP.NET Core project, the wwwroot folder is typically used to store static files like CSS, JavaScript, images, etc. By default, it is not automatically created when you create a new project, but you can manually create it if needed. If your project requ...} (distance percentile=100.0). A domain-specific software answer about ASP.NET project structure.
\item \emph{Yes, sets in C++ are sorted by default. The elements stored in a set are kept in ascending order by default, according to the same strict weak ordering criterion used for the key comparison in map and multimap containers. Under the hood, sets are typically im...} (distance percentile=100.0). A programming-language explanation about C++ sets and ordering.
\item \emph{Medical assistant certification courses can typically take 9 months to 1 year to complete if pursuing a diploma or certificate, and around 2 years if pursuing an associate degree. However, the specific duration may vary based on the educational institution, p...} (distance percentile=100.0). A vocational credentialing answer about medical-assistant certification duration.
\end{itemize}

\subsection*{Low output diversity}
These assistant responses are close to the output-space centroid. They are generic help-oriented greetings or readiness-to-help responses.

\begin{itemize}
\setlength{\itemsep}{3pt}
\setlength{\parsep}{0pt}
\setlength{\parskip}{0pt}
\interlinepenalty=10000
\item \emph{Hello! I'm ChatGPT, an artificial intelligence language model created by OpenAI. I'm here to help answer your questions, provide information, and assist with a wide variety of topics. How can I assist you today?} (distance percentile=0.0). A generic assistant self-introduction and offer to help, close to common output patterns.
\item \emph{I'm here and ready to help you with anything you need! What can I assist you with today?} (distance percentile=0.0). A short generic readiness-to-help response.
\item \emph{Hello! I'm here and ready to help you. What can I assist you with today?} (distance percentile=0.1). A standard greeting and offer of assistance.
\end{itemize}

\subsection*{Searchable inquiry}
These are high-confidence searchable user inquiries, where external web documentation or factual sources could directly answer the request.

\begin{itemize}
\setlength{\itemsep}{3pt}
\setlength{\parsep}{0pt}
\setlength{\parskip}{0pt}
\interlinepenalty=10000
\item \emph{How do I start a simple webserver using nginx?} (classifier confidence=0.9997). A direct how-to query answerable from web documentation.
\item \emph{how do i install mysql with yum on linux} (classifier confidence=0.9997). A software installation query that can be answered from external documentation.
\item \emph{How to apply GPO to only selected group of PCs?} (classifier confidence=0.9997). A system-administration query likely to require searchable Windows/GPO guidance.
\end{itemize}

\subsection*{Non-searchable (unsearchable) inquiry}
These are high-confidence non-searchable user inquiries, where the user asks the model to transform, edit, or inspect provided content rather than retrieve outside information.

\begin{itemize}
\setlength{\itemsep}{3pt}
\setlength{\parsep}{0pt}
\setlength{\parskip}{0pt}
\interlinepenalty=10000
\item \emph{Analyze only the background section and create a slide using the important key elements of the following document: [URL]} (classifier confidence=0.9999). Asks the model to transform supplied material rather than retrieve external facts.
\item \emph{Thank you. This is too difficult for the student. Use simple language and words and keep the length and use various grammatical tenses and structures. Use one idiom please.} (classifier confidence=0.9999). Requests rewriting/simplification of already provided content.
\item \emph{cis this correct robots.txt - Sitemap: [URL] Sitemap: [URL] \# START YOAST BLOCK \# --------------------------- User-agent: * Disallow: Sitemap: [URL] \# --------------------------- \# END YOAST BLOCK} (classifier confidence=0.9999). Asks the model to inspect supplied robots.txt content rather than search externally.
\end{itemize}
\clearpage
\endgroup
\section{OpenAI API Prompts Used to Classify Searchability}
\label{sec:si-prompts}

This section documents the OpenAI API prompt used for the searchability-classification workflow.

\begingroup
\small
\renewcommand{\arraystretch}{1.3}
\noindent\begin{tabular}{@{}p{0.22\linewidth}p{0.74\linewidth}@{}}
\textbf{Analysis use} & Searchability annotation of user questions \\
\textbf{API message role} & \texttt{user} \\
\textbf{Model and parameters} & \texttt{gpt-4o-mini}, temperature \texttt{0.3}, response format JSON object \\
\textbf{Dynamic fields} & \{question\}: raw user query inserted verbatim \\
\end{tabular}
\endgroup
\medskip
\lstdefinestyle{sipromptblock}{
    style=promptblock,
    basicstyle=\ttfamily\fontsize{10}{12}\selectfont,
    frame=single,
    rulecolor=\color{black!22},
    backgroundcolor=\color{black!2},
    framerule=0.4pt,
    framesep=8pt,
    xleftmargin=9pt,
    xrightmargin=9pt,
    aboveskip=8pt,
    belowskip=0pt
}

\begin{lstlisting}[style=sipromptblock]
Please judge whether the following question can obtain a useful answer through Google search.

Question: "{question}"

Please analyze it from the following perspectives:
1. Is this a question for which an answer can be found with a search engine?
2. Why can it, or why can it not, be solved through Google search?

Non-searchable cases include, but are not limited to:
- Pure greetings or small talk (for example, "how are you")
- Questions requiring personalized advice
- Tasks requiring creative generation
- Questions requiring real-time conversation
- Specific code-debugging questions (although related material may be searchable, these usually require personalized analysis)
- Instructions asking the assistant to perform a specific task

Searchable cases include:
- Factual knowledge questions
- General technical questions with standard answers
- Definitions or conceptual explanations
- Historical events and public information

Please respond in JSON format, containing:
{
  "searchable": true/false,
  "confidence": "high/medium/low",
  "reasoning": "brief reason in Chinese or English",
  "category": "question type, such as greeting/factual/creative/debugging/personal"
}
\end{lstlisting}

\clearpage
\section{Supplementary Tables}
\label{sec:si-tables}

\begingroup
\setlength{\parskip}{0pt}
\setlength{\intextsep}{9pt}
\newcommand{\SITableStyle}{%
  \setstretch{1}%
  \fontsize{10}{12}\selectfont
  \renewcommand{\arraystretch}{1.12}%
  \setlength{\tabcolsep}{4pt}%
}
\newcommand{\SITableNote}[1]{%
  \par\vspace{5pt}%
  {\raggedright\fontsize{10}{12}\selectfont\textit{Notes.} #1\par}%
}

\begin{table}[H]
\centering
\SITableStyle
\caption{Sample summary and analysis window.}
\label{tab:s1}
\begin{tabular*}{\linewidth}{@{\extracolsep{\fill}}lr@{}}
\toprule
\textbf{Metric} & \textbf{Value} \\
\midrule
Conversations & 234,839 \\
Conversations with no classified question & 0 \\
\addlinespace[3pt]
User inputs & 442,694 \\
Searchable user inputs & 93,170 \\
Non-searchable user inputs & 349,524 \\
Searchable input share & 0.210461 \\
Non-searchable input share & 0.789539 \\
\addlinespace[3pt]
First-turn searchable conversations & 31,319 \\
First-turn non-searchable conversations & 203,520 \\
First-turn searchable conversation share & 0.133364 \\
First timestamp & 2023-04-09T00:04:52 \\
Last timestamp & 2025-07-31T23:56:04 \\
\bottomrule
\end{tabular*}
\end{table}

\begin{table}[H]
\centering
\SITableStyle
\caption{Searchability classifier confidence and turn-count summaries.}
\label{tab:s2}
\begin{tabular*}{\linewidth}{@{\extracolsep{\fill}}p{0.29\linewidth}rrrrrrrr@{}}
\toprule
\textbf{Group} & \textbf{n} & \textbf{Mean} & \textbf{SD} & \textbf{Median} & \textbf{Min} & \textbf{p25} & \textbf{p75} & \textbf{Max} \\
\midrule
All inputs & 442,694 & 0.987 & 0.058 & 1 & 0.5 & 0.999 & 1 & 1 \\
Searchable inputs & 93,170 & 0.969 & 0.086 & 0.999 & 0.5 & 0.993 & 1 & 1 \\
Non-searchable inputs & 349,524 & 0.991 & 0.046 & 1 & 0.5 & 0.999 & 1 & 1 \\
Turns per conversation & 234,839 & 1.885 & 3.078 & 1 & 1 & 1 & 1 & 350 \\
User questions per conversation & 234,839 & 1.885 & 3.078 & 1 & 1 & 1 & 1 & 350 \\
\bottomrule
\end{tabular*}
\end{table}

\begin{table}[H]
\centering
\SITableStyle
\caption{Weekly searchability trend regressions.}
\label{tab:s3}
\begin{tabular*}{\linewidth}{@{\extracolsep{\fill}}lrrrrl@{}}
\toprule
\multicolumn{6}{@{}l@{}}{\textbf{Turn-level searchable ratio}} \\
\multicolumn{6}{@{}l@{}}{$n =$ 119; $R^2 = 0.757$} \\
\addlinespace[2pt]
\textbf{Term} & \textbf{Estimate} & \textbf{Std. error} & \textbf{$z$ or $t$} & \textbf{$p$ value} & \textbf{95\% CI} \\
\cmidrule(lr){1-6}
const & 0.3286 & 0.0064 & 51.245 & < 1e-300 & [0.3161, 0.3412] \\
week\_index & -0.0022 & 1.24e-04 & -17.454 & 3.22e-68 & [-0.0024, -0.0019] \\
\addlinespace[7pt]
\multicolumn{6}{@{}l@{}}{\textbf{First-turn searchable conversation ratio}} \\
\multicolumn{6}{@{}l@{}}{$n =$ 119; $R^2 = 0.745$} \\
\addlinespace[2pt]
\textbf{Term} & \textbf{Estimate} & \textbf{Std. error} & \textbf{$z$ or $t$} & \textbf{$p$ value} & \textbf{95\% CI} \\
\cmidrule(lr){1-6}
const & 0.2465 & 0.0063 & 39.004 & < 1e-300 & [0.2341, 0.2589] \\
week\_index & -0.0019 & 1.05e-04 & -18.306 & 7.41e-75 & [-0.0021, -0.0017] \\
\bottomrule
\end{tabular*}
\SITableNote{Weighted by denominator count; HC1 robust SE.}
\end{table}

\clearpage
\setlength{\intextsep}{6pt}

\begin{table}[H]
\centering
\SITableStyle
\caption{Conversation-level response knowledge-extent summaries.}
\label{tab:s4}
\begin{tabular*}{\linewidth}{@{\extracolsep{\fill}}lrrrrrrr@{}}
\toprule
\textbf{Measure} & \textbf{Mean} & \textbf{SD} & \textbf{Median} & \textbf{Min} & \textbf{p25} & \textbf{p75} & \textbf{Max} \\
\midrule
\multicolumn{8}{@{}l@{}}{\textbf{overall} \quad $n =$ 229,043} \\
ke\_max & 0.157 & 0.297 & 0 & 0 & 0 & 0 & 1.326 \\
ke\_mean & 0.135 & 0.249 & 0 & 0 & 0 & 0 & 0.957 \\
ke\_median & 0.131 & 0.243 & 0 & 0 & 0 & 0 & 0.958 \\
ke\_std & 0.014 & 0.038 & 0 & 0 & 0 & 0 & 0.347 \\
\addlinespace[5pt]
\multicolumn{8}{@{}l@{}}{\textbf{non\_searchable} \quad $n =$ 207,194} \\
ke\_max & 0.125 & 0.267 & 0 & 0 & 0 & 0 & 1.326 \\
ke\_mean & 0.108 & 0.226 & 0 & 0 & 0 & 0 & 0.95 \\
ke\_median & 0.105 & 0.22 & 0 & 0 & 0 & 0 & 0.955 \\
ke\_std & 0.011 & 0.034 & 0 & 0 & 0 & 0 & 0.333 \\
\addlinespace[5pt]
\multicolumn{8}{@{}l@{}}{\textbf{searchable} \quad $n =$ 43,688} \\
ke\_max & 0.242 & 0.329 & 0 & 0 & 0 & 0.525 & 1.257 \\
ke\_mean & 0.215 & 0.285 & 0 & 0 & 0 & 0.489 & 0.937 \\
ke\_median & 0.211 & 0.28 & 0 & 0 & 0 & 0.478 & 0.946 \\
ke\_std & 0.018 & 0.039 & 0 & 0 & 0 & 0 & 0.302 \\
\bottomrule
\end{tabular*}
\end{table}

\begin{table}[H]
\centering
\SITableStyle
\caption{Coverage of the paired searchable/non-searchable response knowledge-extent analysis.}
\label{tab:s5}
\begin{tabular*}{\linewidth}{@{}p{\linewidth}@{}}
\toprule
\textbf{Analysis coverage} \\
\midrule
\raggedright The paired analysis includes 21,839 conversations with both searchable and non-searchable response knowledge-extent measurements. The paired-difference test is reported in Table~\ref{tab:s7}. The conversation-level records are not included in this source package.\tabularnewline
\bottomrule
\end{tabular*}
\end{table}

\begin{table}[H]
\centering
\SITableStyle
\caption{Response-count summaries.}
\label{tab:s6}
\begin{tabular*}{\linewidth}{@{\extracolsep{\fill}}lrrrrrr@{}}
\toprule
\textbf{Group} & \textbf{Count} & \textbf{Mean} & \textbf{Std} & \textbf{Median} & \textbf{Min} & \textbf{Max} \\
\midrule
non\_searchable & 234,839 & 1.463 & 2.434 & 1 & 0 & 350 \\
overall & 234,839 & 1.857 & 3.081 & 1 & 0 & 350 \\
searchable & 234,839 & 0.395 & 1.493 & 0 & 0 & 168 \\
\bottomrule
\end{tabular*}
\end{table}

\begin{table}[H]
\centering
\SITableStyle
\caption{Paired searchable/non-searchable response knowledge-extent difference test.}
\label{tab:s7}
\begin{tabular*}{\linewidth}{@{\extracolsep{\fill}}lrrrrl@{}}
\toprule
\multicolumn{6}{@{}l@{}}{\textbf{Paired non-searchable minus searchable KE\_mean}} \\
\multicolumn{6}{@{}l@{}}{$n =$ 21,839} \\
\addlinespace[2pt]
\textbf{Term} & \textbf{Estimate} & \textbf{Std. error} & \textbf{$z$ or $t$} & \textbf{$p$ value} & \textbf{95\% CI} \\
\cmidrule(lr){1-6}
const & 0.082 & 0.003 & 31.836 & 2.09e-222 & [0.0767, 0.0867] \\
\bottomrule
\end{tabular*}
\SITableNote{Intercept-only paired-difference test; HC1 SE.}
\end{table}

\clearpage
\setlength{\intextsep}{9pt}

\begin{table}[H]
\centering
\SITableStyle
\renewcommand{\arraystretch}{1.45}
\caption{Next-turn diversity regressions and robustness checks.}
\label{tab:s8}
\begin{tabular*}{\linewidth}{@{\extracolsep{\fill}}lrrrrl@{}}
\toprule
\multicolumn{6}{@{}l@{}}{\textbf{Main: log next-user diversity, standardized predictors}} \\
\multicolumn{6}{@{}l@{}}{$n =$ 30,000; $R^2 = 0.294$} \\
\addlinespace[2pt]
\textbf{Term} & \textbf{Estimate} & \textbf{Std. error} & \textbf{$z$ or $t$} & \textbf{$p$ value} & \textbf{95\% CI} \\
\cmidrule(lr){1-6}
const & -0.024 & 1.39e-04 & -172.816 & < 1e-300 & [-0.0243, -0.0237] \\
ai\_t\_diversity\_std & 0.005 & 1.67e-04 & 29.652 & 3.20e-193 & [0.0046, 0.0053] \\
user\_t\_diversity\_std & 0.011 & 1.97e-04 & 56.023 & < 1e-300 & [0.0107, 0.0115] \\
\addlinespace[15pt]
\multicolumn{6}{@{}l@{}}{\textbf{Robustness: level-level}} \\
\multicolumn{6}{@{}l@{}}{$n =$ 30,000; $R^2 = 0.295$} \\
\addlinespace[2pt]
\textbf{Term} & \textbf{Estimate} & \textbf{Std. error} & \textbf{$z$ or $t$} & \textbf{$p$ value} & \textbf{95\% CI} \\
\cmidrule(lr){1-6}
const & 0.345 & 0.007 & 50.373 & < 1e-300 & [0.3318, 0.3586] \\
ai\_t\_diversity & 0.223 & 0.007 & 29.909 & 1.50e-196 & [0.2083, 0.2375] \\
user\_t\_diversity & 0.422 & 0.007 & 56.249 & < 1e-300 & [0.4072, 0.4365] \\
\addlinespace[15pt]
\multicolumn{6}{@{}l@{}}{\textbf{Robustness: log-log}} \\
\multicolumn{6}{@{}l@{}}{$n =$ 30,000; $R^2 = 0.293$} \\
\addlinespace[2pt]
\textbf{Term} & \textbf{Estimate} & \textbf{Std. error} & \textbf{$z$ or $t$} & \textbf{$p$ value} & \textbf{95\% CI} \\
\cmidrule(lr){1-6}
const & -0.01 & 1.96e-04 & -52.668 & < 1e-300 & [-0.0107, -0.0099] \\
log\_ai\_t\_diversity & 0.224 & 0.008 & 29.836 & 1.33e-195 & [0.2097, 0.2392] \\
log\_user\_t\_diversity & 0.421 & 0.008 & 55.629 & < 1e-300 & [0.4058, 0.4354] \\
\bottomrule
\end{tabular*}
\SITableNote{Cluster-robust SE by conversation.}
\end{table}
\clearpage

\begin{table}[H]
\centering
\SITableStyle
\renewcommand{\arraystretch}{1.45}
\ContinuedFloat
\caption[]{Next-turn diversity regressions and robustness checks. (continued)}
\begin{tabular*}{\linewidth}{@{\extracolsep{\fill}}lrrrrl@{}}
\toprule
\multicolumn{6}{@{}l@{}}{\textbf{Robustness: log(x+0.001)}} \\
\multicolumn{6}{@{}l@{}}{$n =$ 30,000; $R^2 = 0.293$} \\
\addlinespace[2pt]
\textbf{Term} & \textbf{Estimate} & \textbf{Std. error} & \textbf{$z$ or $t$} & \textbf{$p$ value} & \textbf{95\% CI} \\
\cmidrule(lr){1-6}
const & -0.01 & 1.91e-04 & -52.141 & < 1e-300 & [-0.0103, -0.0096] \\
log\_ai\_t\_diversity & 0.224 & 0.008 & 29.836 & 1.33e-195 & [0.2097, 0.2392] \\
log\_user\_t\_diversity & 0.421 & 0.008 & 55.63 & < 1e-300 & [0.4058, 0.4354] \\
\addlinespace[15pt]
\multicolumn{6}{@{}l@{}}{\textbf{Robustness: excluding near-zero diversity}} \\
\multicolumn{6}{@{}l@{}}{$n =$ 30,000; $R^2 = 0.295$} \\
\addlinespace[2pt]
\textbf{Term} & \textbf{Estimate} & \textbf{Std. error} & \textbf{$z$ or $t$} & \textbf{$p$ value} & \textbf{95\% CI} \\
\cmidrule(lr){1-6}
const & 0.345 & 0.007 & 50.373 & < 1e-300 & [0.3318, 0.3586] \\
ai\_t\_diversity & 0.223 & 0.007 & 29.909 & 1.50e-196 & [0.2083, 0.2375] \\
user\_t\_diversity & 0.422 & 0.007 & 56.249 & < 1e-300 & [0.4072, 0.4365] \\
\addlinespace[15pt]
\multicolumn{6}{@{}l@{}}{\textbf{Robustness: winsorized 1/99}} \\
\multicolumn{6}{@{}l@{}}{$n =$ 30,000; $R^2 = 0.296$} \\
\addlinespace[2pt]
\textbf{Term} & \textbf{Estimate} & \textbf{Std. error} & \textbf{$z$ or $t$} & \textbf{$p$ value} & \textbf{95\% CI} \\
\cmidrule(lr){1-6}
const & 0.345 & 0.007 & 50.829 & < 1e-300 & [0.3312, 0.3578] \\
ai\_t\_diversity & 0.223 & 0.007 & 29.914 & 1.31e-196 & [0.208, 0.2372] \\
user\_t\_diversity & 0.423 & 0.007 & 56.652 & < 1e-300 & [0.4082, 0.4375] \\
\bottomrule
\end{tabular*}
\SITableNote{Cluster-robust SE by conversation.}
\end{table}

\endgroup

\clearpage
\endgroup
\end{document}